\documentclass[journal]{IEEEtran}

\usepackage{ifpdf}
\usepackage{authblk}

\usepackage{cite}

\ifCLASSINFOpdf
  \usepackage[pdftex]{graphicx}
\else
  \usepackage[dvips]{graphicx}
\fi

\usepackage{amsmath,bm}
\usepackage{amsfonts}
\usepackage{amssymb}
\usepackage{color,soul}
\usepackage{soul,color}
\soulregister\cite7
\soulregister\ref7

\ifCLASSOPTIONcompsoc
\usepackage[caption=false,font=normalsize,labelfont=sf,textfont=sf]{subfig}
\else
   \usepackage[caption=false,font=footnotesize]{subfig}
\fi

\usepackage{algorithmic}

\usepackage{array}

\usepackage{fixltx2e}
\newcommand\blfootnote[1]{%
  \begingroup
  \renewcommand\thefootnote{}\footnote{#1}%
  \addtocounter{footnote}{-1}%
  \endgroup
}

\begin{document}
\title{Semi-Sequential Probabilistic Model For Indoor Localization Enhancement}
\author{Minh Tu Hoang, Brosnan Yuen, Xiaodai Dong, Tao Lu, Robert Westendorp, and Kishore Reddy}
\maketitle

%\boldmath
\begin{abstract}
\blfootnote{
This work was supported in part by the Natural Sciences and Engineering Research Council of Canada under Grant 520198, Fortinet Research under Contract 05484 and NVidia under GPU Grant program (\textit{Corresponding authors: X. Dong and T. Lu.}). \\
M. T. Hoang, B.Yuen, X. Dong and T. Lu are with the
Department of Electrical and Computer Engineering, University of Victoria,
Victoria, BC, Canada (email: \{xdong, taolu\}@ece.uvic.ca).\\
R. Westendorp and K. Reddy are with Fortinet Canada Inc., Burnaby, BC,
Canada.}
This paper proposes a semi-sequential probabilistic model (SSP) that applies an additional short term memory to enhance the performance of the probabilistic indoor localization. The conventional probabilistic methods normally treat the locations in the database indiscriminately. In contrast, SSP leverages the information of the previous position to determine the probable location since the user's speed in an indoor environment is bounded and locations near the previous one have higher probability than the other locations. Although the SSP utilizes the previous location information, it does not require the exact moving speed and direction of the user. On-site experiments using the received signal strength indicator (RSSI) and channel state information (CSI) fingerprints for localization demonstrate that SSP reduces the maximum error and boosts the performance of existing probabilistic approaches by 25$\%$ - 30$\%$. 

\textit{Index Terms}- Received signal strength indicator (RSSI), channel state information (CSI), WiFi indoor localization, probabilistic approach based localization. 
% no keywords
\end{abstract}

%%%%%%%%%%%%%%%%%%%%%%%%%%%%%%%%%%%%%%%%%%%%%%%%%%%%%%%%%%%%%%%%%%%%%%%%%%%%%%%%%%%%%%%%%%%%%%%%%%%%%%%%%%%%%%%
\section{Introduction} \label{sec:intro} 
%%%%%%%%%%%%%%%%%%%%%%%%%%%%%%%%%%%%%%%%%%%%%%%%%%%%%%%%%%%%%%%%%%%%%%%%%%%%%%%%%%%%%%%%%%%%%%%%%%%%%%%%%%%%%%%%
Indoor localization using Wi-Fi fingerprints enables a wide variety of applications~\cite{He2016}. For example, customers in a shopping mall can use navigation and tracking services to explore  stores and to find their desired destinations~\cite{Zafari2019}. In a museum, accurate indoor localization can transform a visitor's phone into a virtual guide for contextual information based on his/her location. Among all the available solutions to indoor localization, Wi-Fi fingerprinting~\cite{Xu2016} is one of the most popular ones due to the wide availability of Wi-Fi devices, which eliminates the requirement for additional infrastructure and hardware. There are two popular types of Wi-Fi fingerprints: received signal strength indicator (RSSI) and channel state information (CSI). RSSI provides the signal power received at the receiver and can be retrieved from most Wi-Fi receivers such as mobile phones, tablets, laptops, etc.~\cite{Chen2016} RSSI has two drawbacks: instability due to fading and multipath~\cite{Haochen2017} and the lack of information. %In addition, RSSI requires at least 3 routers for triangulation. 
On the other hand, CSI has higher granularity than the RSSI as it can capture both the amplitude and phase responses of the channel in different frequencies and between separate transmitter-receiver antennae pairs~\cite{Zafari2019}. However, CSI is only available with specific wireless network interface cards (NIC) such as the Intel WiFi Link 5300 MIMO NIC, Atheros AR9390, and Atheros AR9580 chipset~\cite{Wang2017b}. Several methods have been proposed to utilize RSSI, CSI or both for accurate indoor localization. 

Fingerprinting based Wi-Fi localization can be realized by deterministic, probabilistic or machine learning approaches~\cite{He2016}. The deterministic method, e.g., K nearest neighbors (KNN)~\cite{Bahl2000,YaqinXie2016}, uses a similarity metric to identify the measured signal from the fingerprint database and estimate the user's position. Despite its low complexity, the accuracy of KNN is unstable due to the substantial fluctuations of Wi-Fi signals~\cite{YaqinXie2016,DongLi2016,Brunato2005}. On the other hand, machine learning algorithms such as multilayer perceptron (MLP)~\cite{Battiti2002}, robust extreme learning machine (RELM)~\cite{Lu2016}, multi-layer neural network (MLNN)~\cite{Dai2016}, convolutional neural network (CNN)~\cite{Jiao2017}, etc., provide highest accuracy~\cite{C.Gentile2013,Laoudias2009} while most of them are sophisticated in nature and require extremely high computational complexity in training~\cite{Brunato2005}. In contrast, probabilistic methods are based on the statistical inference between the measured signal  and the stored fingerprints through Bayes rule~\cite{Youssef2005}. In general, they have medium complexity and provide good accuracy in indoor localization~\cite{Xu2016, Zafari2019}. Therefore, probabilistic methods are widely utilized to extract information from both RSSI and CSI fingerprints.  

In order to determine the location, probabilistic methods require the probability density function (PDF) of the fingerprinting features. Some early research assume the RSSI PDF follows empirical parametric distributions such as  single-peak~\cite{Kaemarungsi2004} or double-peak Gaussian~\cite{L.Chen2013}, lognormal distribution~\cite{Honkavirta2009}, etc. Ref.~\cite{Kaemarungsi2004} observes experimentally that most RSSI values follow single-peak Gaussian distribution and tend to be left-skewed. The left-skewed distribution occurs when the variation of the weaker RSSI is larger than that of the stronger one. In a later research, Ref.~\cite{Honkavirta2009} indicates that the RSSI distribution can be left-skewed, symmetric or right-skewed depending on the distance and obstacles between the user and access points (AP). Therefore, the lognormal PDF is more suitable to approximate the right- or left-skewed RSS distribution. Although a common assumption about the RSS distribution is single-peak Gaussian, Ref.~\cite{L.Chen2013} claims that RSSI can be double-peak Gaussian distribution (DGD) under some circumstances and suggested to replace the single-peak Gaussian distribution with DGD. However, all of these observations are highly dependant on the experiment specifics and may not be  reasonable approximations~\cite{QidengJiang2016}.         

To achieve better performance by eliminating the assumption on the RSSI PDF, \cite{Roos2002} exploits both histogram and Kernel methods to estimate the PDF. In contrast, such methods are called non-parametric. Here, the histogram of RSSI is estimated for each AP with a set of non-overlapping bins that cover the whole range of the RSSIs for each AP. Then the RSSI PDF is calculated as a piecewise constant function where the density is constant within each bin. In~\cite{Kushki2007}, the experiment by Kushki \textit{et al.} shows that a bin width of 10 dB provides the lowest positioning error. Works in Horus~\cite{Youssef2005} compare the performance between parametric and  histogram methods. It is found that both of those methods provide comparable performances with a slight advantage for the parametric one. The reason is from the existence of zero count bins in the histogram due to the limited number of different signal strength values in the training phase. As an improvement to the histogram method, Kernel method employs component smoothing functions for each data value to produce a smooth and continuous probability curve and avoid any zero count bins. Ref.~\cite{C.Figuera2009} further proposes a non-parametric statistical model with Parzen window density estimation. The kernel for Parzen window needs to be non-negative and normalized. Among all of the suitable kernels such as Epanechnikov, logistic and sigmoid, etc., Gaussian kernel is claimed to have consistently good performance and is the most widely used.

In order to construct a fingerprint map for accurate localization, a large number of reference points (RP) are required~\cite{Jun2018}, which is time-consuming and labor-intensive~\cite{He2016}. Recently, probabilistic Gaussian process (GP) is utilized to enhance the accuracy of indoor localization in the uncalibrated domain with a limited number of RPs. GP is another non-parametric model characterized completely by its mean function and co-variance matrix. Ref.~\cite{Sun2018} presents GP regression models to predict the spatial distribution of RSSI. The appropriate compound kernel functions are systematically selected instead of a single kernel function to get the heterogeneous RSSI PDF. In Ref.~\cite{Yiu2016}, GP is trained by the firefly algorithm (FA) to obtain the hyper-parameters. Once the GP hyper-parameters are obtained, the GP prior distribution can be used for regression to predict RSSI at locations with no prior measurements.

Besides RSSI, CSI fingerprints are widely used in probabilistic methods. FILA~\cite{Wu2013}, one of the early works using CSI, estimates the signal strength distribution for each AP at each location based on the total power of all CSI sub-carriers. After the CSI power distribution database is constructed, the probabilistic method with Bayes' rule predicts the user's location. Later research on BiLoc~\cite{Wang2017b} exploits bi-modal data that estimates the angle of arrival (AoA) and average amplitudes of CSI as the fingerprints for localization. The advantage of the method is that when there is no line of sight (LoS), the CSI amplitude will be significantly reduced but AoA will be less affected. In contrast, when LoS is available, the CSI amplitude is a stable fingerprint to be relied on. In the testing phase, the probabilistic approach is adopted to localize the user's position using those bi-modal fingerprints.

\begin{table*}[!t]
\centering         
\caption{Comparisons of Indoor Localization Experiments Using Probabilistic Techniques} 
\label{table:AverageErr} 
% \begin{adjustbox}{width=1\textwidth} 
\begin{tabular}{l c c c c c c c c} 
\hline           
\textbf{Method} & \textbf{Feature} & \textbf{Access point (AP)} & \textbf{Reference point (RP)} & \textbf{Testing Point} & \textbf{Grid Size}  & \textbf{Accuracy}\\ 
DGD~\cite{L.Chen2013} & RSSI  &  - &  68 &  35  & 2.5 m & 2.8 m\\
Horus~\cite{Youssef2005} & RSSI  &  21 &  172 &  100  & 1.52 m & 0.42 $\pm$ 0.28 m\\
Kernel method\cite{Kushki2007} & RSSI  &  33 &  66 &  44  & 2 m & 2.31 $\pm$ 2.10 m\\
BiLoc \cite{Wang2017b} & CSI  &  1 &  15 &  15  & 1.8 m & 1.5 $\pm$ 0.8 m\\
FILA~\cite{Wu2013} & CSI   &  1-3 &  28 &  -  & - & 0.4 m to 1 m\\
\hline         
\end{tabular} 
% \end{adjustbox}
\end{table*}

Table~\ref{table:AverageErr} summarizes the experiment specifics and  results of typical probabilistic methods. Here the number of APs, RPs and testing points vary among experiments and the grid size is defined as the distance between two consecutive RPs. 

In general, probabilistic methods provide acceptable accuracy from 1~m to 2.5~m. However, all of the above methods treat all the locations in the database with equal probability in predicting the current location, which ignores its correlation to the user's previous position. Since the moving speed of the user in an indoor environment is bounded, the locations near the previous one should have higher probability to be estimated as the user's current point than others. Therefore, in this paper, we propose a simple short term memory step for all existing probabilistic methods to enhance their performances. Our semi-sequential probabilistic model (SSP) applies window functions such as Gaussian, Hann and Tukey, and is based on the physical distance between the RP and the user's predicted previous position to calculate the probability of RP being near the user's current position. As a result, the spatial ambiguity of fingerprints~\cite{Minh2018} is significantly reduced and the localization accuracy is improved. 

In the literature, the idea of exploiting the measurements in previous time steps to locate the current location was adopted in the research of recurrent neural network (RNN)~\cite{Minh2019}, Kalman filter~\cite{Au2013, Guvenc2003, Besada2007, Kushki2006} and soft range limited K-nearest neighbors (SRL-KNN)~\cite{Minh2018}. In Ref.~\cite{Minh2019}, the proposed P-MIMO LSTM model exploits the sequential RSSI measurements and the trajectory information from multiple time steps to achieve high accuracy. However, the complexity is high, and the long term memory dependency can cause the significant accumulated errors if the inaccuracy in historical data is high. On the other hand, Kalman filter~\cite{Au2013} estimates the most likely current location based on prior measurements and Gaussian noises with linear motion dynamics assumptions. In real scenarios, those assumptions are not necessarily valid~\cite{YogitaChapre2013}. SRL-KNN~\cite{Minh2018} does not require the above assumptions and reaches the lowest complexity. However, the modified penalty functions in SRL-KNN can only be applied for Euclidean distance, not for probabilistic model. In contrast to those approaches, SSP is able to boost the performance of several probabilistic systems using  RSSI or CSI fingerprints. Furthermore, the short term memory dependency ensures our low complexity and avoid accumulating errors. The proposed model is tested with several experiments using both RSSI and CSI fingerprints and compared with existing probabilistic methods.          
        
%%%%%%%%%%%%%%%%%%%%%%%%%%%%%%%%%%%%%%%%%%%%%%%%%%%%%%%%%%%%%%%%%%%%%%%%%%%%%%%%%%%%%%%%%%%%%%%%%%%%%%%%%%%%%%%
\section{Proposed Model} \label{sec:model} 
%%%%%%%%%%%%%%%%%%%%%%%%%%%%%%%%%%%%%%%%%%%%%%%%%%%%%%%%%%%%%%%%%%%%%%%%%%%%%%%%%%%%%%%%%%%%%%%%%%%%%%%%%%%%%%%
\subsection{Proposed Localization System} \label{sec:Notation}
The proposed localization system has two phases: a training phase and a testing phase. In the training phase, fingerprints at each predefined reference point (RP) location are collected and stored in a database. The fingerprints can be either RSSI or CSI. We assume the area of interest has $P$ APs and $M$ RPs. For each RP $i$ at its physical location $\bm{l}_{i}(x_{i},y_{i})$, a corresponding fingerprint vector is denoted as $\bm{F}(\bm{l}_i)=\{F_{1}(\bm{l}_i), F_{2}(\bm{l}_i),...,F_{N}(\bm{l}_i)\}$, where $N$ is the number of available features and $F_{j}(\bm{l}_i),\, 1 \leq j \leq N,$ is the $j$-th feature at point $i$. In the testing phase, each unknown location of the user, denoted as a testing point, is determined by the localization algorithm. 

%%%%%%%%%%%%%%%%%%%%%%%%%%%%%%%%%%%%%%%%%%%%%%%%%%%%%%%%%%%%%%%%%%%%%%%%%%%%%%%%%%%%%%%%%%%%%%%%%%%%%%%%%%%%%%%
\subsection{Proposed Semi-Sequential Probabilistic (SSP) Model}
%%%%%%%%%%%%%%%%%%%%%%%%%%%%%%%%%%%%%%%%%%%%%%%%%%%%%%%%%%%%%%%%%%%%%%%%%%%%%%%%%%%%%%%%%%%%%%%%%%%%%%%%%%%%%%%
\subsubsection{Conventional Probabilistic Method} \label{subsec:Conventional}
In the testing phase, we assume that a fingerprint vector $\bm{F}(\bm{l}_{curr})$ at the unknown current location $\bm{l}_{curr}$ is measured. The likelihood function $P(F_{k}(\bm{l}_{curr})|\bm{l}_{i})$ describes the probability of the $k$-th fingerprint feature at location $\bm{l}_i$, has a signal strength $F_k(\bm{l}_i)\approx{F_{k}}(\bm{l}_{curr})$ condition on $\bm{l}_{curr}\approx\bm{l}_i$, which can be estimated by the parametric~\cite{Kaemarungsi2004,L.Chen2013,Honkavirta2009} or non-parametric~\cite{Roos2002,Youssef2005} methods. Therefore, the probability of the current location $\bm{l}_{curr}$ being close to $\bm{l}_i$ given the measured fingerprint vector $\bm{F}(\bm{l}_{curr})$ can be estimated according to Bayes' theorem~\cite{Xu2016}
\begin{equation}  \label{prob1}
P(\bm{l}_{curr}\approx\bm{l}_{i}|\bm{F}(\bm{l}_{curr}))= \prod_{k=1}^{N} \frac{P(F_{k}(\bm{l}_{curr})|\bm{l}_{i})P(\bm{l}_{i})}{P(F_{k}(\bm{l}_{curr}))}
\end{equation}
Here, Eq.~\eqref{prob1} is valid following the assumptions along with other researchers that $F_k$ are mutually independent, which leads to $P(\bm{F}(\bm{l}_{curr})|\bm{l}_i)=\prod_{k=1}^{N}P(F_k(\bm{l}_{curr})|\bm{l}_{i})$  and $P(\bm{F}(\bm{l}_{curr}))=\prod_{k=1}^{N}P(F_k(\bm{l}_{curr}))$. Each location $\bm{l}_{i}$ has a priori probability $P(\bm{l}_{i}){\equiv}P(\bm{l}_{i}{\approx}\bm{l}_{curr})$ to be nearest to $\bm{l}_{curr}$, which is initially assumed to be equally likely for every location in the database. Furthermore, $\prod_{k=1}^{N} P(F_{k}(\bm{l}_{curr}))$ is independent of $\bm{l}_{i}$. Therefore, for simplification, Eq.~\eqref{prob1} can be modified to

\begin{equation}  \label{prob2}
P(\bm{l}_{curr} \approx \bm{l}_{i}|\bm{F}(\bm{l}_{curr})) \propto \prod_{k=1}^{N} P(F_{k}(\bm{l}_{curr})|\bm{l}_{i})
\end{equation} 

Eq.~\eqref{prob2} is applied for the whole database to get the set of $P(\bm{l}_{curr}\approx\bm{l}_{i}|\bm{F}(\bm{l}_{curr}))$ including all of the RPs. The maximum value of $P(\bm{l}_{curr}\approx\bm{l}_{i}|\bm{F}(\bm{l}_{curr}))$ can be chosen as the user's location~\cite{Youssef2005}. In some approaches~\cite{L.Chen2013}, $K$ biggest values are chosen, and the user's location is the average of all $K$ values.

\subsubsection{Proposed SSP Model} \label{sec:ssp} 
\begin{figure}[!t]
     \centering
\subfloat[\label{fig:Circular}]
{\includegraphics[width=0.45\textwidth]{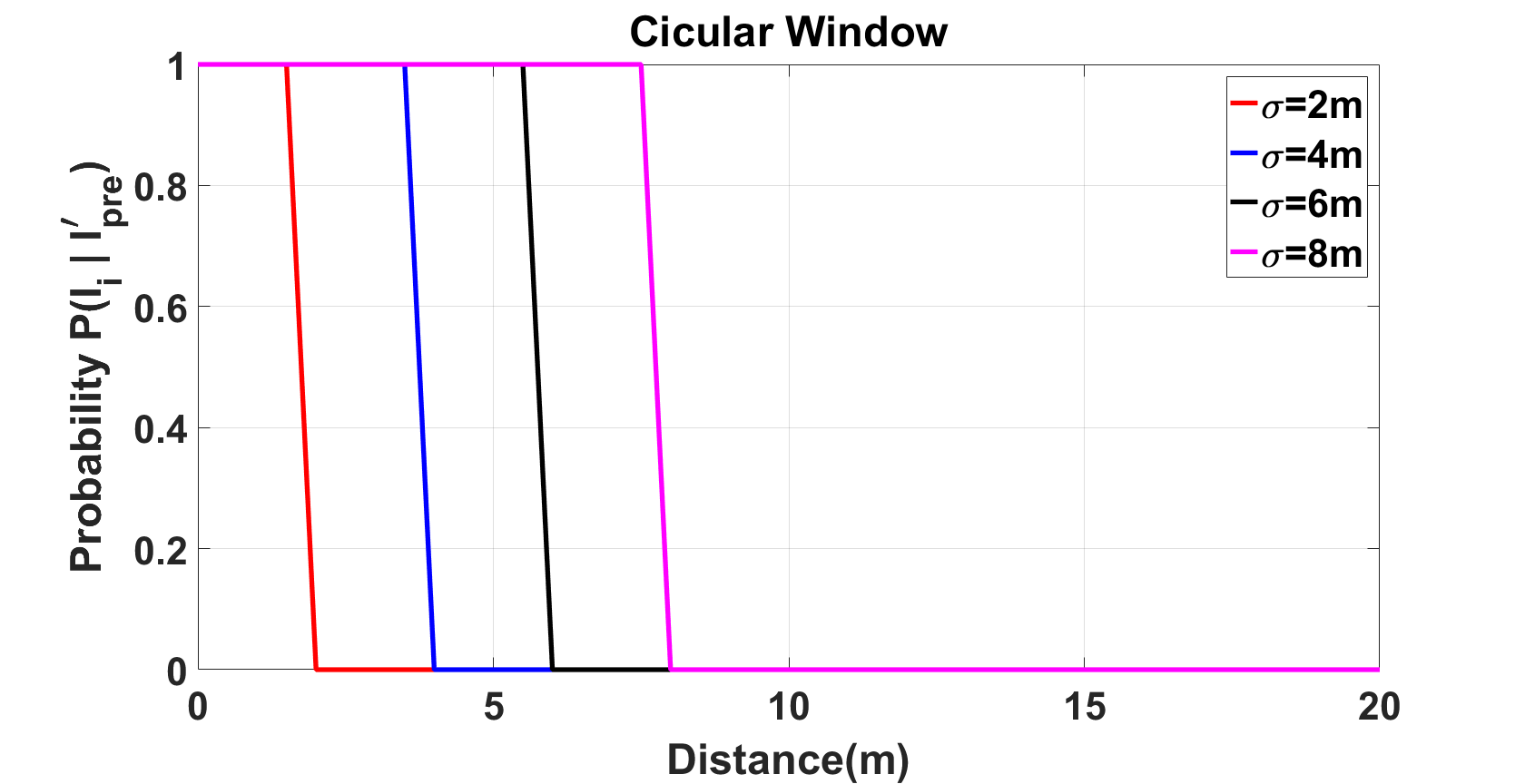}}\quad
\subfloat[\label{fig:Gaussian}]{\includegraphics[width=0.45\textwidth]{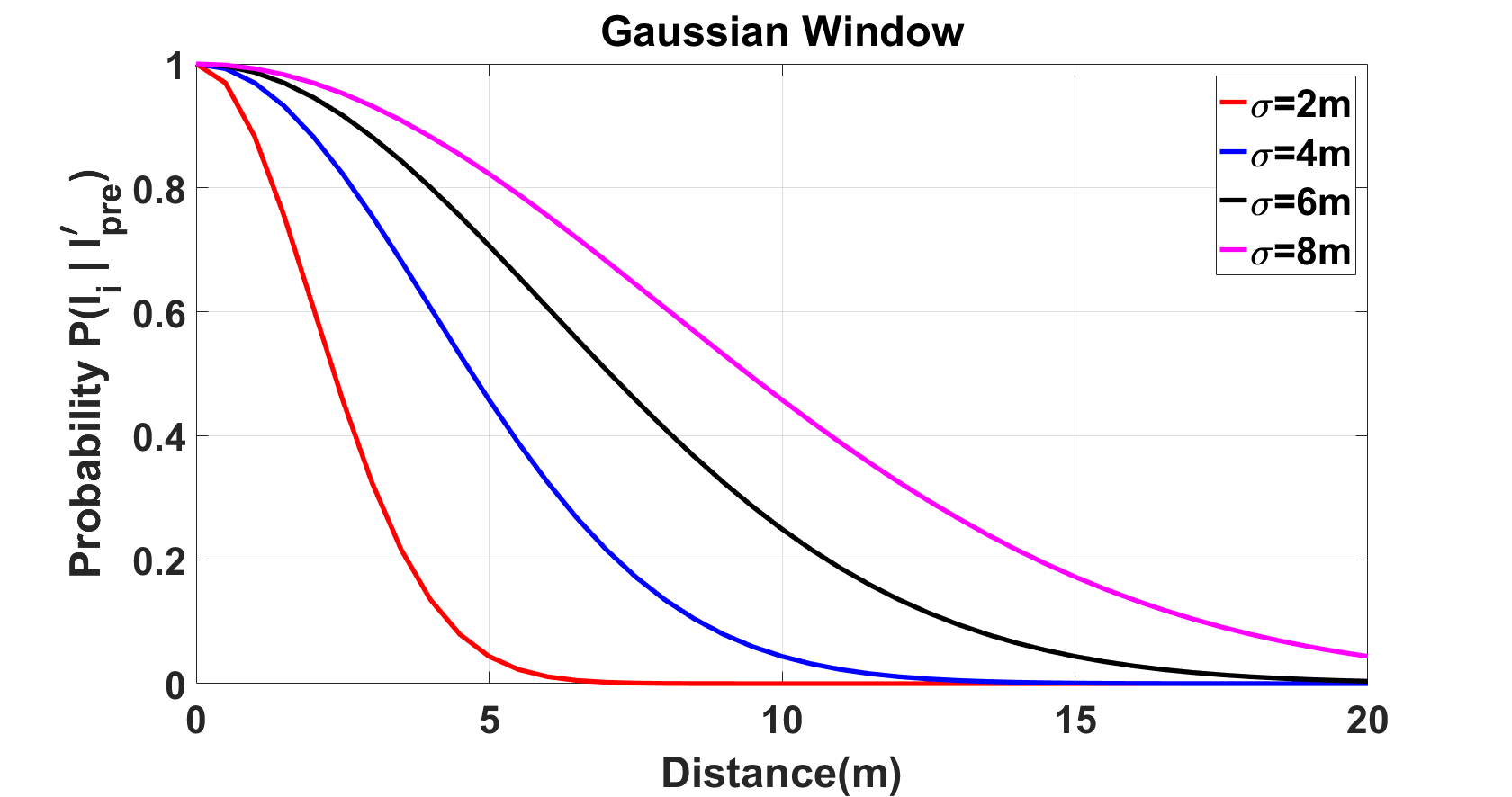}}\quad
\subfloat[\label{fig:Hann}]{\includegraphics[width=0.45\textwidth]{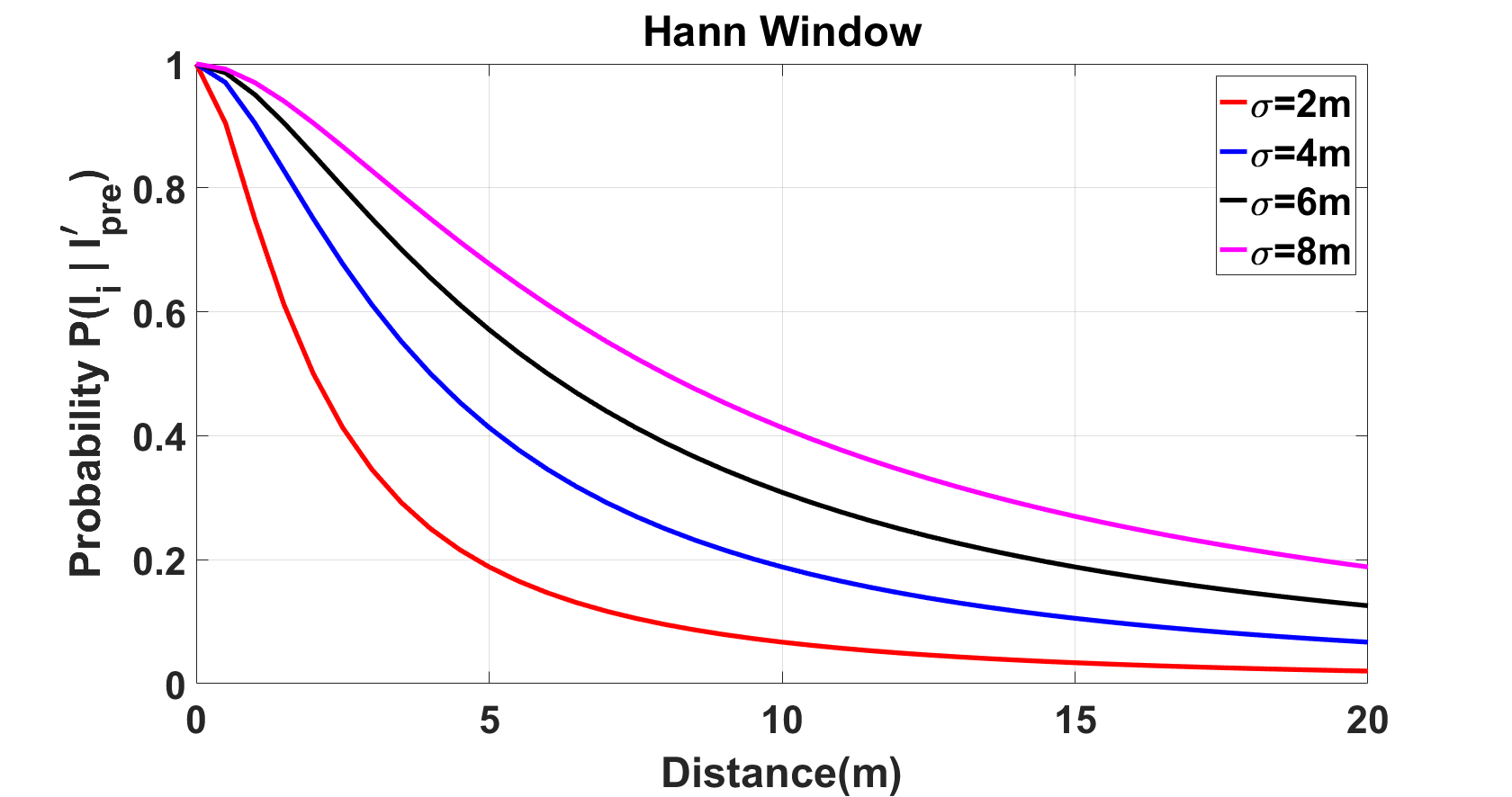}}\quad
\subfloat[\label{fig:Tukey}]{\includegraphics[width=0.45\textwidth]{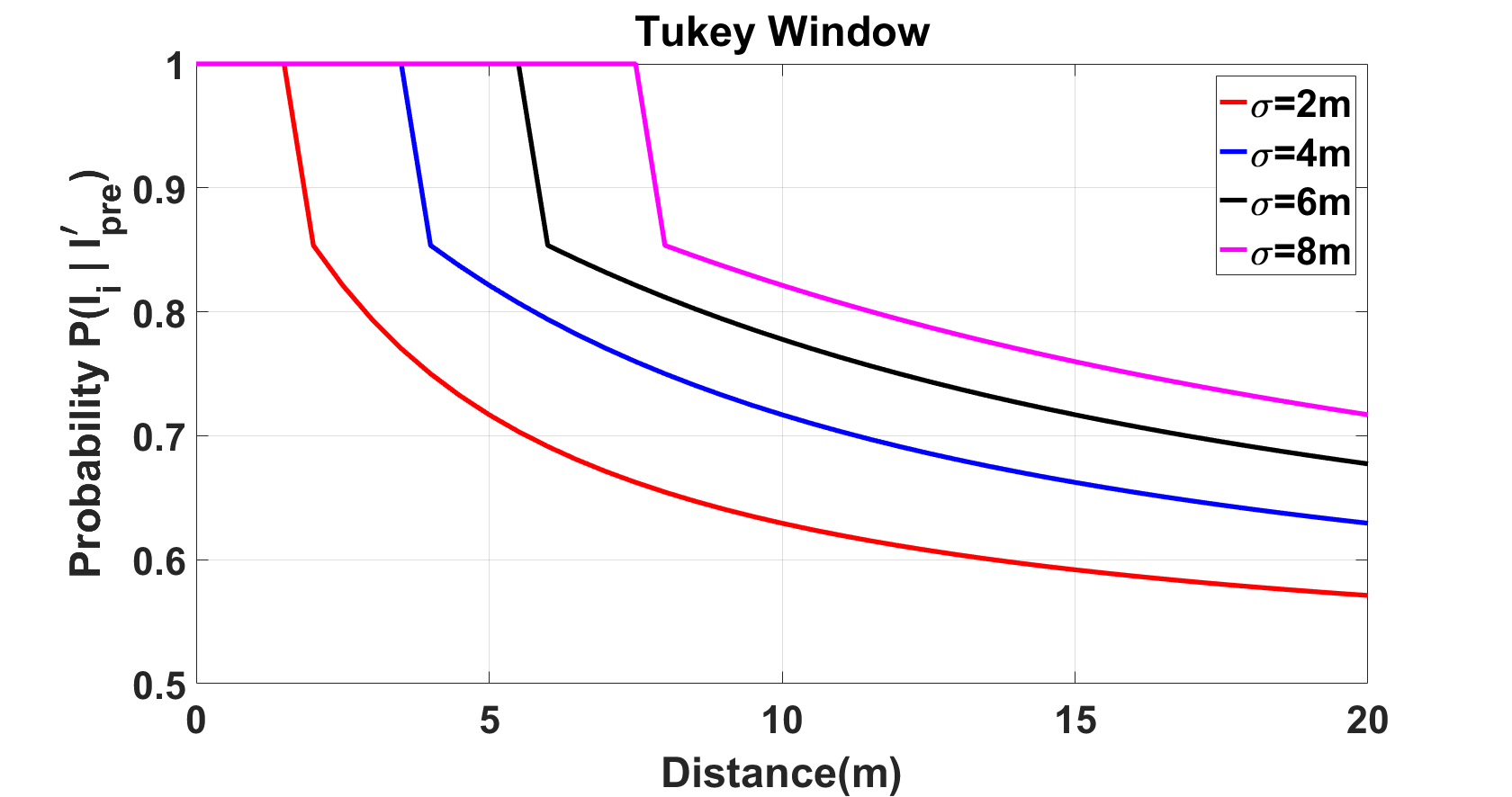}}
     \caption{Short term memory window forms with different values of $\sigma$. (a) Circular Window. (b)  Gaussian Window. (c) Hann Window. (d) Tukey Window.}
     \label{fig:window_form}
\end{figure}

As the moving speed of a user is bounded, it is highly unlikely for a user to move to an unrealistic distant position from the prior one during the consecutive measurements. Therefore, the assumption that $P(\bm{l}_{i})$ is equally likely for every location is not valid.  In fact, the locations near the previous one should have higher probability than the others. In order to incorporate that into the probabilistic models, a short term memory window is proposed to determine the possibility of a reference location $\bm{l}_{i}$ to be the nearest to the current location $\bm{l}_{curr}$. For example, a circular window whose radius is determined by the user moving speed and time duration between two consecutive measurements can be drawn around the previous location to limit the nearest neighbor search within the area. Besides the hard range limit, we here propose several soft range windows based on the predicted previous location $\bm{l}^{\prime}_{pre}(x_{pre}^\prime,y_{pre}^\prime)$:
\begin{itemize}
    \item[--] Gaussian window:
    \begin{equation} \label{Gauss}
    P(\bm{l}_{i}|\bm{l}^{\prime}_{pre}) = \exp(-\frac{D^2}{2\sigma^2})
    %}{\sum_{i=1}^{M} P(\bm{l}_{i})}
    \end{equation}  
    
    \item[--] Hann window:
    \begin{equation} \label{Hann}
     P(\bm{l}_{i}|\bm{l}^{\prime}_{pre}) = \cos^{2}\left[{\frac{\pi D}{2\cdot(D+\sigma)}}\right]
     \end{equation} 
     
     \item[--] Tukey window:
    \begin{equation} \label{Tukey}
     P(\bm{l}_{i}|\bm{l}^{\prime}_{pre}) =  
     \begin{cases}
        1, & \text{if } D<\sigma\\
       \frac{1}{2}\left\{1+\cos\left[\frac{\pi{D}}{2\cdot(D+\sigma)}\right]\right\}, & \text{if } D \geq \sigma
    \end{cases}   
     \end{equation}  
\end{itemize}
where $D(\bm{l}_i,\bm{l}_{pre}^\prime)=\sqrt{(x_{i}-x^{\prime}_{pre})^2+(y_{i}-y^{\prime}_{pre})^2}$ is the distance between $\bm{l}_{i}$ and  $\bm{l}^{\prime}_{pre}$.  The standard deviation $\sigma$ determines the spread of the window. Fig.~\ref{fig:window_form} illustrates the shapes of the above windows and the dependency between the probability of the considered location in the database $P(\bm{l}_{i}|\bm{l}^{\prime}_{pre})$ and the distance $D$. Clearly, $P(\bm{l}_{i}|\bm{l}^{\prime}_{pre})$ is inversely proportional to $D$. Here, $\sigma$ can be determined according to the maximum distance $d_{max}$ that a user can move during the sampling time interval $\Delta{t}$. As shown in Fig.~\ref{fig:window_form}, the larger the $\sigma$ is, the more spread out the short term memory window becomes.   

After applying the short term memory window for the whole database, the probability $P(\bm{l}_{i}|\bm{l}^{\prime}_{pre})$ can be normalized as follows.

\begin{equation} \label{P_l}
\hat{P}(\bm{l}_{i}|\bm{l}^{\prime}_{pre}) = \frac{P(\bm{l}_{i}|\bm{l}^{\prime}_{pre})}{\sum_{i=1}^{M} P(\bm{l}_{i}|\bm{l}^{\prime}_{pre})}
 \end{equation} 

For simplicity, we denote $\hat{P}(\bm{l}_{i}|\bm{l}^{\prime}_{pre})$  as $P(\bm{l}_{i}|\bm{l}^{\prime}_{pre})$ in the rest of the paper unless otherwise specified. Finally, Eq.~\eqref{prob1} is modified with the above short term memory step as follows.

\begin{equation}  \label{prob5_1}
\begin{aligned}
P(\bm{l}_{curr}\approx\bm{l}_{i}|\bm{F}(\bm{l}_{curr}),\bm{l}^{\prime}_{pre})= \\
\prod_{k=1}^{N} \frac{P(F_{k}(\bm{l}_{curr})|\bm{l}_{i},\bm{l}^{\prime}_{pre})P(\bm{l}_{i}|\bm{l}^{\prime}_{pre})}{P(F_{k}(\bm{l}_{curr}))}
\end{aligned}
\end{equation} 
$\prod_{k=1}^{N} P(F_{k}(\bm{l}_{curr}))$ can still be ignored here due to the same argument in Subsection \ref{subsec:Conventional}. As $P(F_{k}(\bm{l}_{curr})|\bm{l}_{i},\bm{l}^{\prime}_{pre})$ is independent of the previous predicted location $\bm{l}^{\prime}_{pre}$, i.e. $P(F_{k}(\bm{l}_{curr})|\bm{l}_{i},\bm{l}^{\prime}_{pre}) = P(F_{k}(\bm{l}_{curr})|\bm{l}_{i})$, Eq.~\eqref{prob5_1} becomes

\begin{equation}  \label{prob5}
P(\bm{l}_{curr}\approx\bm{l}_{i}|\bm{F}(\bm{l}_{curr}),\bm{l}^{\prime}_{pre})\propto \\
\prod_{k=1}^{N} P(F_{k}(\bm{l}_{curr})|\bm{l}_{i})P(\bm{l}_{i}|\bm{l}^{\prime}_{pre})
\end{equation}    

 With this simple short term memory, we can convert the existing memoryless probabilistic methods, e.g., Horus~\cite{Youssef2005}, DGD~\cite{L.Chen2013}, Kernel method~\cite{Kushki2007}, FILA~\cite{Wu2013}, BiLoc~\cite{Wang2017b}, to semi-sequential model and improve their performances significantly.  

%%%%%%%%%%%%%%%%%%%%%%%%%%%%%%%%%%%%%%%%%%%%%%%%%%%%%%%%%%%%%%%%%%%%%%%%%%%%%%%%%%%%%%%%%%%%%%%%%%%%%%%%%%%%%%%%%
\section{Database And Experiments} \label{sec:experiment}

\begin{figure*}[!t]
\centering
\includegraphics[width=\textwidth]{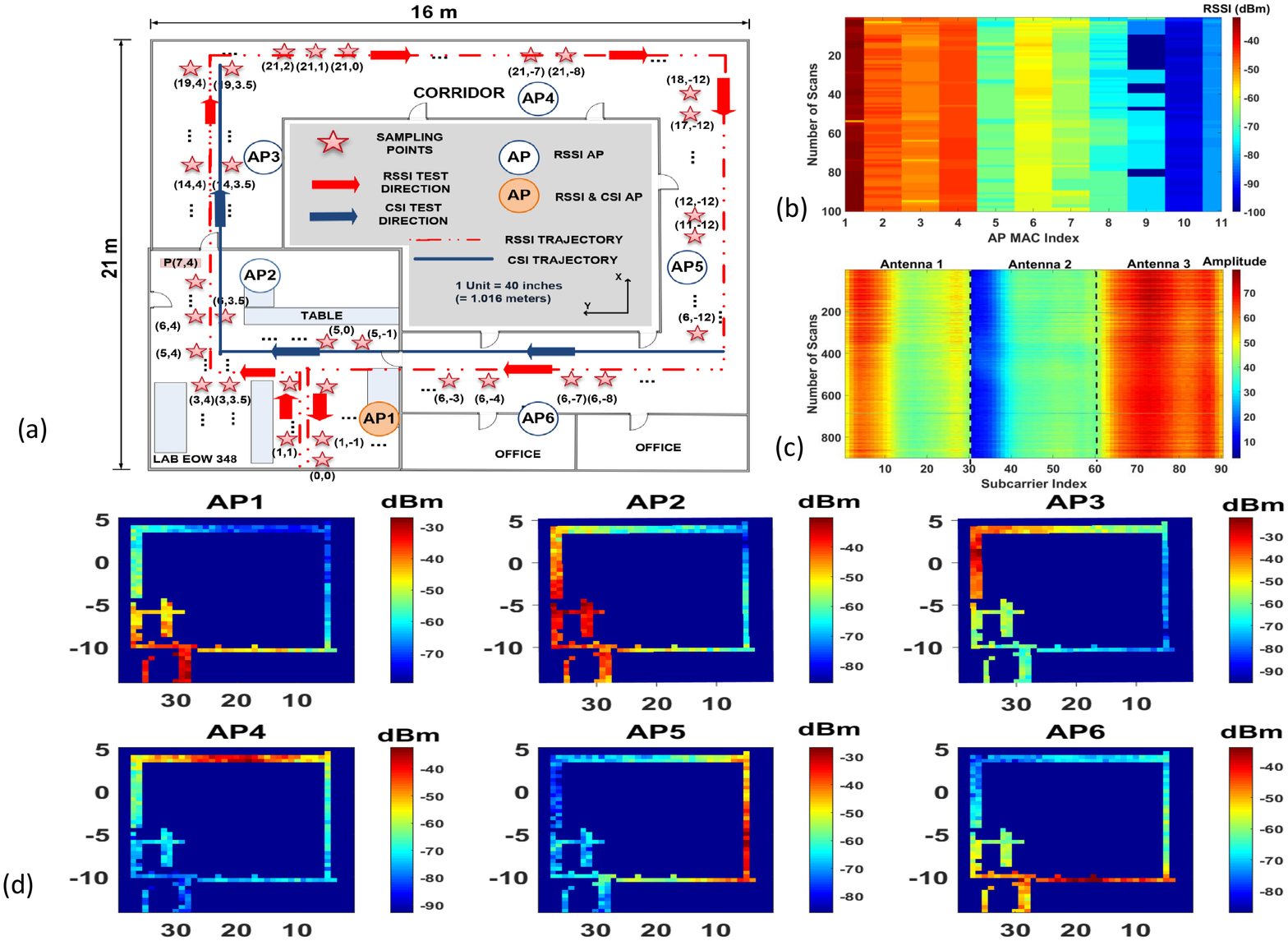}
\caption{(a) Floor map of the RSSI and CSI test site. The solid red line is the RSSI trajectory with red arrows pointing toward walking direction. The dash green line is the CSI trajectory. (b) An example of a collected RSSI vector with 100 scans. (c) An example of a collected CSI image from Intel 5300 NIC. (d) Heat map of the RSSI strength from 6 APs used in our localization scheme.}
\label{fig:floor_map}
\end{figure*}

This paper investigates both RSSI and CSI fingerprints for localization. All experiments have been carried out on the third floor of Engineering Office Wing (EOW), University of Victoria, BC, Canada. The dimension of the area is 21~m by 16~m. It has three long corridors as shown in Fig.~\ref{fig:floor_map}(a). The RSSI and CSI fingerprints for both training and testing are collected using an autonomous driving robot. The 3-wheel robot has multiple sensors including a wheel odometer, an inertial measurement unit (IMU), a LIDAR, 3 sonar sensors and a color and depth (RGB-D) camera. It can navigate to a target location within an accuracy of $\mathrm{0.07{\pm}0.02~m}$.

During RSSI localization experiments, the robot carried a mobile device (Google Nexus 4 running Android 4.4) to collect RSSI. In the training phase, 100 RSSI scans ($S_1=100$) were obtained at each location. Fig.~\ref{fig:floor_map}(b) illustrates an instance of the 100-scan RSSI from 11 AP network interface controllers (NICs). In this experiment, there are 6 APs and 5 of them have 2 distinct MAC address each assigned to two associating NICs with 2.4~GHz and 5~GHz communications channels respectively, except for one that operates on 2.4~GHz frequency only. In total, each scan contains 11 RSSI readings. In the testing phase, the user location will be updated at a consecutive sampling time interval $\Delta{t} = 1$~s. During that time interval, only a small number of RSSI scans ($S_{2}=$1 or 2) were available for the fingerprint matching. Fig.~\ref{fig:floor_map}(a) illustrates our localization scheme with $365$ RPs (pink stars) and $175$ testing points (dash red line). Fig.~\ref{fig:floor_map}(d) shows the heat map of $6$ APs, where the signal strength is represented by color. Clearly, the signals from $6$ APs cover the whole targeting area including $1$ room and $3$ corridors.   

In the same targeting area, only one single AP (AP1) that has 3 receiving antennas with 30 multiple subcarriers each, was set up for CSI measurements. The robot carrying an Intel WiFi Link 5300 MIMO NIC navigated across 145~RPs and 60~testing points (along the blue solid line in Fig.~\ref{fig:floor_map}(a)) to collect CSI with csitool~\cite{Halperin2011}. A single CSI data is a complex number to represent its amplitude and phase. However, the phase of CSI is noisy due to effects such as fading and frequency fluctuation~\cite{Haochen2017,Wang2017}. Therefore, only CSI amplitude is utilized in this paper. In each measurement at RP $i$, we group $H$ number of CSI measurements with $W$ subcarriers  to construct an $H \times W$ matrix 
\begin{equation} \label{eq:am_csi}
\centering   
{\tilde A}(\bm{l}_{i}) = 
\begin{bmatrix}
    {\tilde A}^{11}_i       & {\tilde A}^{12}_i  & \ldots & {\tilde A}^{1W}_i \\
    {\tilde A}^{21}_i       & {\tilde A}^{22}_i  & \ldots & {\tilde A}^{2W}_i \\
    \vdots          & \vdots    & \ddots & \vdots \\
    {\tilde A}^{H1}_i       & {\tilde A}^{H2}_i  & \ldots & {\tilde A}^{HW}_i \\
\end{bmatrix}.
\end{equation} 
Here, ${\tilde A}^{hw}_i$ is the CSI amplitude value of the subcarrier $w$ in the measurement $h$ at RP $i$. Fig.~\ref{fig:floor_map}(c) shows an image representation of collected CSI with $H=900$ and $W=90$. 

In the testing phase, the robot moved along a pre-defined route (Fig.~\ref{fig:floor_map}(a)) at a speed  randomly changing within (0.6-4.0)~m/s to simulate the indoor walking pattern of a normal person~\cite{Browning2006,Email2007}. The user location was updated every $\Delta{t}=$1~s. The testing experiments are repeated several times in different days and time.

%%%%%%%%%%%%%%%%%%%%%%%%%%%%%%%%%%%%%%%%%%%%%%%%%%%%%%%%%%%%%%%%%%%%%%%%%%%%%%%%%%%%%%%%%%%%%%%%%%%%%%%%%%%%%%%%%                   %%%%%%%%%%%%%%%%%%%%%%%%%%%%%%%%%%%%%%%%%%%%%%%%%%%%%%%%%%%%%%%%%%%%%%%%%%%%%%%%%%%%%%%%%%%%%%%%%%%%%%%%%%%%%
\section{Results And Discussions} \label{sec:sim_result}
\subsection{SSP and Conventional Methods Comparison}

\begin{table}[!t]
\centering         
\caption{Average Errors (SSP Model with Gaussian Window, $\sigma=d_{max}$)} 
\label{table:SSP} 
% \begin{adjustbox}{width=1\textwidth} 
\begin{tabular}{l c c c c} 
\hline           
\textbf{Method} & \textbf{Fingerprint} & \textbf{AP Number} & \textbf{Original Model (m)} & \textbf{SSP Model (m)}\\ 
Horus & RSSI & 6 &$1.5\pm 1.2$ & $1.0\pm0.7$\\
DGD & RSSI & 6 &$1.5\pm1.3$ & $1.0\pm0.8$\\
Kernel  & RSSI & 6 & $1.3\pm1.0$ & $1.0\pm0.7$\\
FILA & CSI & 1 &$4.4\pm2.4$ & $2.2\pm1.4$\\
Biloc & CSI & 1 &$4.8\pm2.3$ & $2.5\pm2.0$\\
\hline         
\end{tabular} 
% \end{adjustbox}
\end{table}

The proposed SSP is added to memoryless probabilistic methods including Horus~\cite{Youssef2005}, DGD~\cite{L.Chen2013}, Kernel method~\cite{Kushki2007}, BiLoc~\cite{Wang2017b} and FILA~\cite{Wu2013}. As the maximum speed, $v_{max}$ is pre-configured to be 4~m/s, the maximum distance a user can travel during consecutive measurements is $d_{max}=v_{max}\Delta{t}=$4~m. Therefore, we used a Gaussian window with $\sigma=d_{max}$ in the SSP model. 

Among all RSSI models, Horus~\cite{Youssef2005} and DGD~\cite{L.Chen2013} are typical parametric approaches with the assumption of a single-peak Gaussian distributed $P(\bm{F}(\bm{l}_{curr})|\bm{l}_{i})$. The Kernel method~\cite{Kushki2007} is a non-parametric model which builds the probability distribution by using the kernel smoothing function. The additional short term memory step is added to calculate $P(\bm{l}_{i})$ following Subsection~\ref{sec:ssp}.    

Among CSI models, FILA~\cite{Wu2013} exploits CSI amplitude to build a single-peak Gaussian distributed database $P(\bm{F}(\bm{l}_{curr})|\bm{l}_{i})$. A possible approach to calculate $P(\bm{l}_{i})$ is  to use Pearson correlation between the testing and the stored CSI fingerprints in the database. However, this approach relies only on the spatial information, which ignores the time related information from the previous location. Instead, SSP model determines $P(\bm{l}_{i})$ based on the user's previous location which adds the valuable information from time domain. On the other hand, BiLoc~\cite{Wang2017b} exploits bi-modal data as the fingerprints. During training, a deep autoencoder network with a pre-train restricted Boltzmann machine (RBM) is used to extract the unique channel features. In the testing, following the radial basis function (RBF), a Bayesian probability model is employed to estimate position. The probability $P(\bm{F}(\bm{l}_{curr})|\bm{l}_{i})$ is measured according to the difference between the original and reconstructed CSI. The probability $P(\bm{l}_{i})$ is assumed to be uniformly distributed. In contrast, SSP model adds an additional short term memory step to BiLoc similar to RSSI cases.

Table~\ref{table:SSP} illustrates the average localization errors of probabilistic algorithms before and after applying SSP. The chosen parameters of SSP are Gaussian window (Eq.~\eqref{Gauss}) with $\sigma = d_{max}$. Clearly, with the additional sequential consideration, the proposed SSP models show the significant performance improvement comparing with the original approaches. For Horus, the average error decreases $33 \%$ from $\mathrm{1.5{\pm}1.2~m}$ to as low as $\mathrm{1.0{\pm}0.7~m}$ by applying SSP. Similarly, SSP boosts the performance of the DGD and Kernel method by ${\sim}33\%$ and ${\sim}25\%$ respectively. In the cases of CSI fingerprinting localization, SSP helps to improve the accuracy of FILA by ${\sim}50\%$, from $\mathrm{4.4{\pm}2.3~m}$ to $\mathrm{2.2{\pm}1.4~m}$. Among all models, the most significant improvement is achieved in the case of BiLoc with $48\%$ reduction of the average error from $\mathrm{4.8{\pm}2.3~m}$ to $\mathrm{2.5{\pm}2.0~m}$.  

\begin{figure}[!t]
\centering
\includegraphics[width=0.5\textwidth]{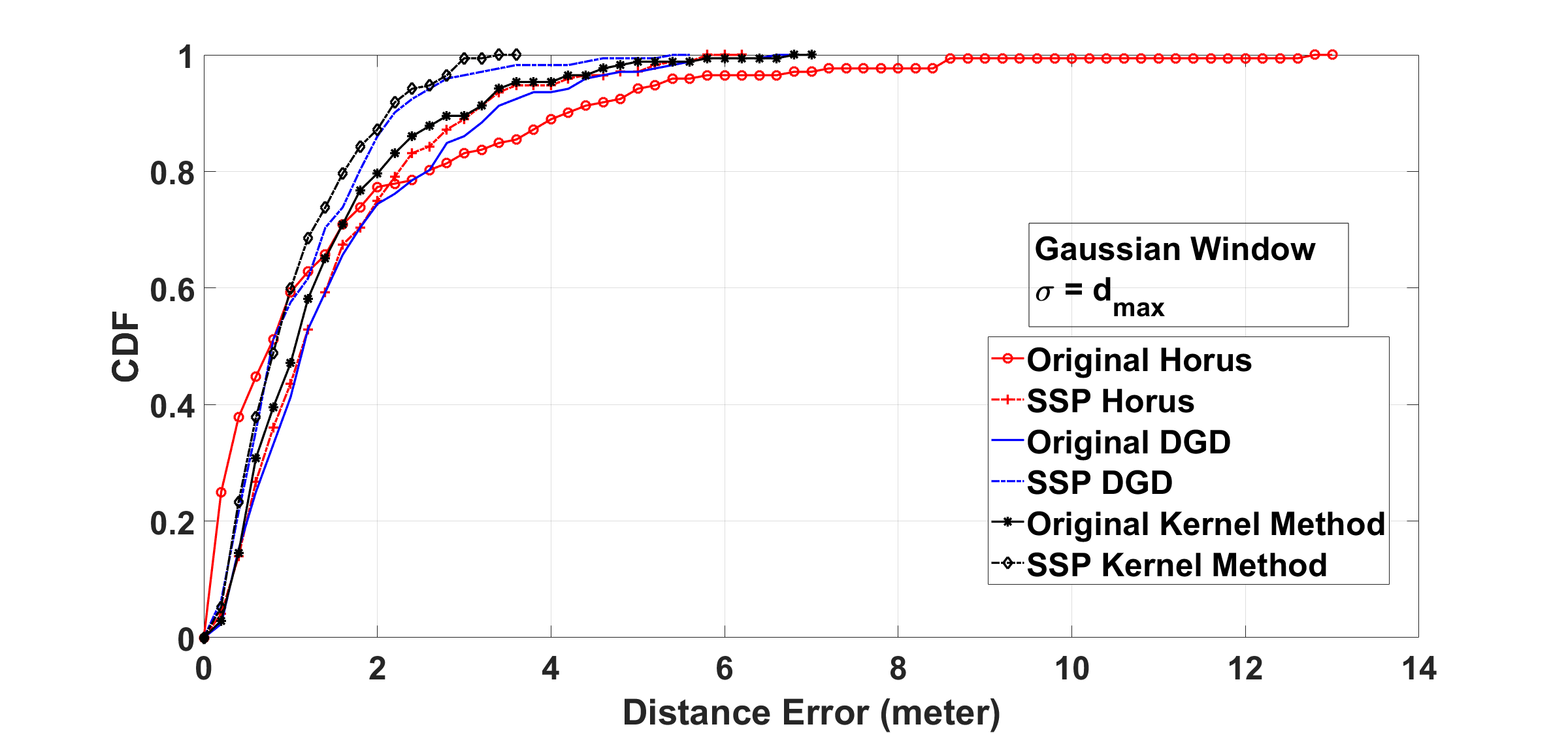}
\caption{CDF of the localization error of conventional probabilistic models and SSP models with RSSI fingerprint.}
\label{fig:RSSI_sigma_4}
\end{figure}

\begin{figure}[!t]
\centering
\includegraphics[width=0.5\textwidth]{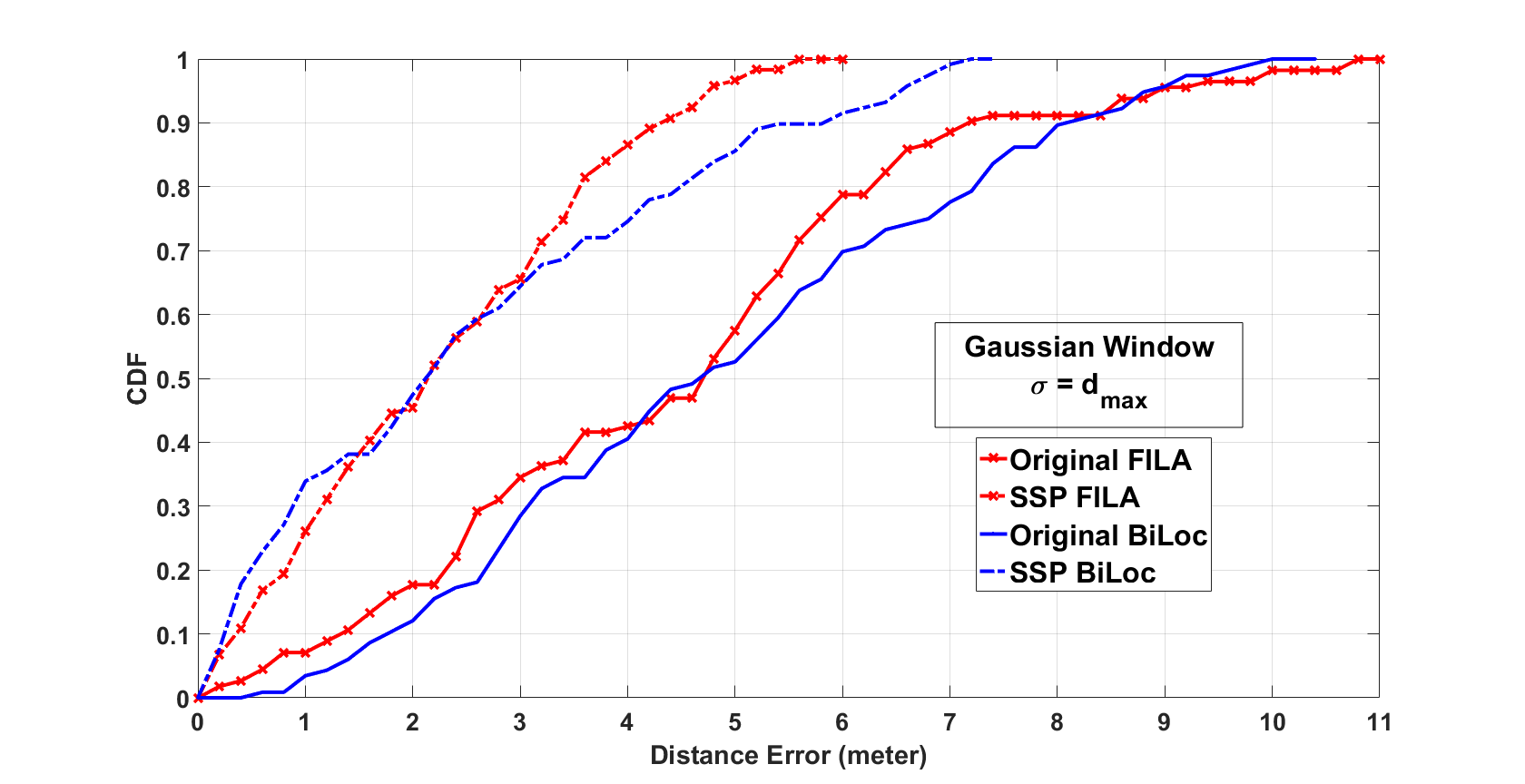}
\caption{CDF of the localization error of conventional probabilistic models and SSP models with CSI fingerprint.}
\label{fig:CSI_sigma_4}
\end{figure}

In addition to the reduction of average error, SSP also reduces  the maximum error of all conventional methods significantly. Fig.~\ref{fig:RSSI_sigma_4} compares the cumulative distribution function (CDF) of localization errors of Horus, DGD and Kernel method before and after adopting SSP.  Due to larger RSSI fluctuations, the memoryless approaches may choose a wrong location with similar fingerprints, which could be far from the actual location, leading to an extreme large error in the scale of the testing site dimension. As shown in Fig.~\ref{fig:RSSI_sigma_4}, without SSP, the maximum error can be as high as 13~m for Horus and 7~m for DGD and Kernel method. In contrast, SSP eliminates such error pattern with the short term memory step, resulting in a much smaller maximum error of 6~m for Horus, 5.5~m for DGD and as low as 3.5~m for Kernel method. Besides, as shown in Fig.~\ref{fig:CSI_sigma_4}, for CSI experiment with 1 router, SSP helps to reduce the maximum error of FILA from 11~m to below 6~m and BiLoc from 10~m to $\sim$7~m.

Evidently, the above results show that SSP can significantly boost the performance of both RSSI and CSI fingerprints based probabilistic algorithms. Besides, as shown in Table~\ref{table:SSP}, the localization accuracy of the RSSI experiments with 6 APs is significantly better than that of CSI experiments with 1 AP, i.e., around $1$ m of SSP models based RSSI compared with more than $2$ m of SSP models based CSI. Therefore, in the following sections, to analyze  important parameters such as window types and the corresponding $\sigma$ value, we focus on the RSSI experiment.
%%%%%%%%%%%%%%%%%%%%%%%%%%%%%%%%%%%%%%%%%%%%%%%%%%%%%%%%%%%%%%%%%%%%%%%%%%%%%%%%%%%%%%%%%%%%%%%%%%%%%%%%%%%%%%%
\subsection{Parameter Analysis}
\subsubsection{Window Type}
\begin{figure*}[!t]
\centering
\includegraphics[width=\textwidth]{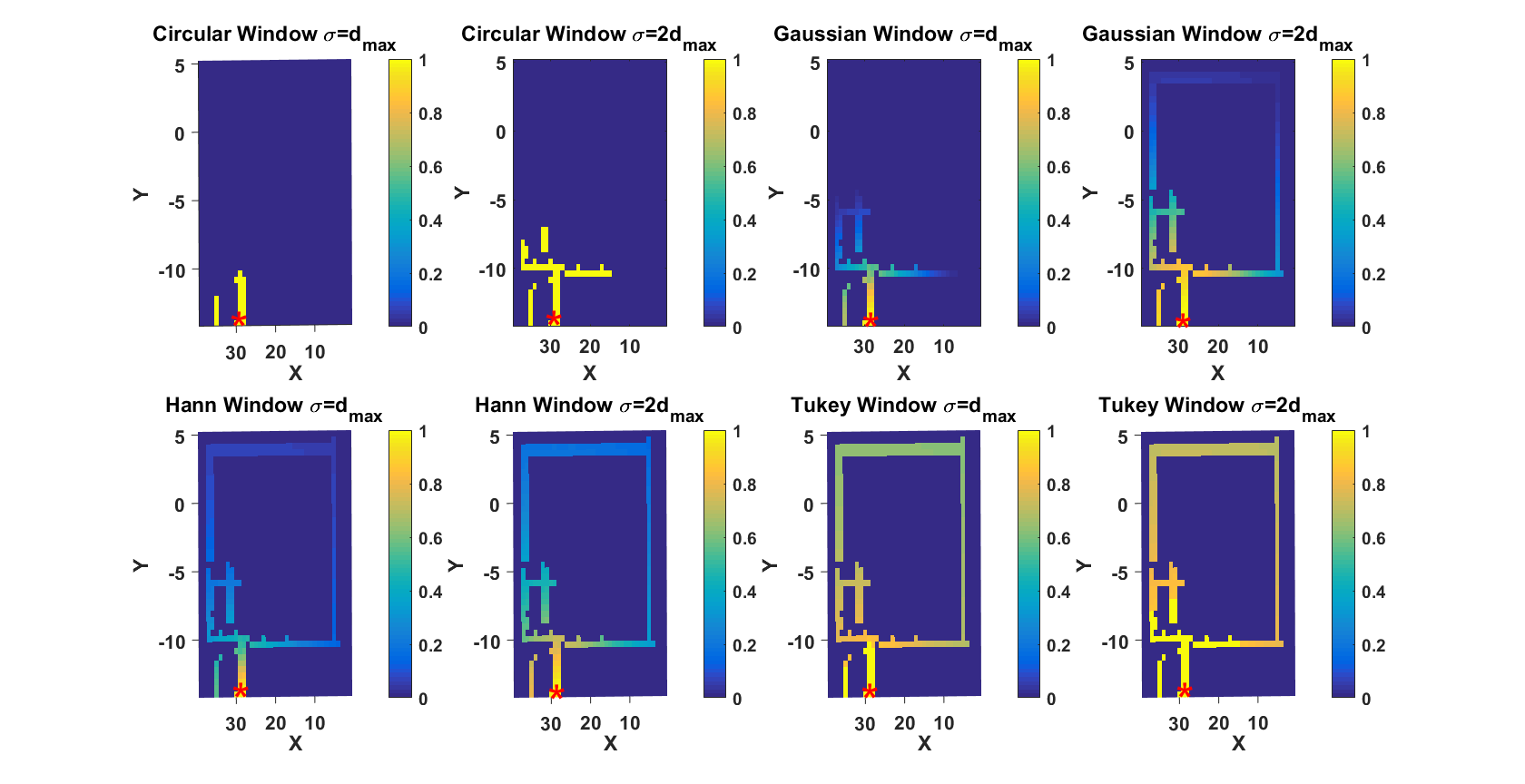}
\caption{Probabilistic heat map of all RPs in the database after applying SSP windows (the red star represents for the user's previous location).}
\label{fig:prob_heatmap}
\end{figure*}

Following discussions in Subsection~\ref{sec:ssp}, we propose three more soft range SSP windows including Gaussian, Hann and Tukey in addition to the simple hard range circular window. Fig.~\ref{fig:window_form} illustrates the shapes of these windows. Among all, the circular window has a clear drop at the circumference $D=\sigma$, which eliminates the possibility of any RPs having the distances bigger than $\sigma$ being considered as close to the current localization. Gaussian and Hann windows have the smooth decline curves with the increase of distance $D$. When $\sigma$ increases, these curves spread out more. On the other hand, Tukey window is the combination between hard range and soft range forms with the equivalent highest probability for all RPs with distance smaller than $\sigma$ and the gradual dropping curve following the increase of distance $D$.

\begin{figure}[!t]
     \centering
\subfloat[\label{fig:Diff_Circular}]
{\includegraphics[width=0.48\textwidth]{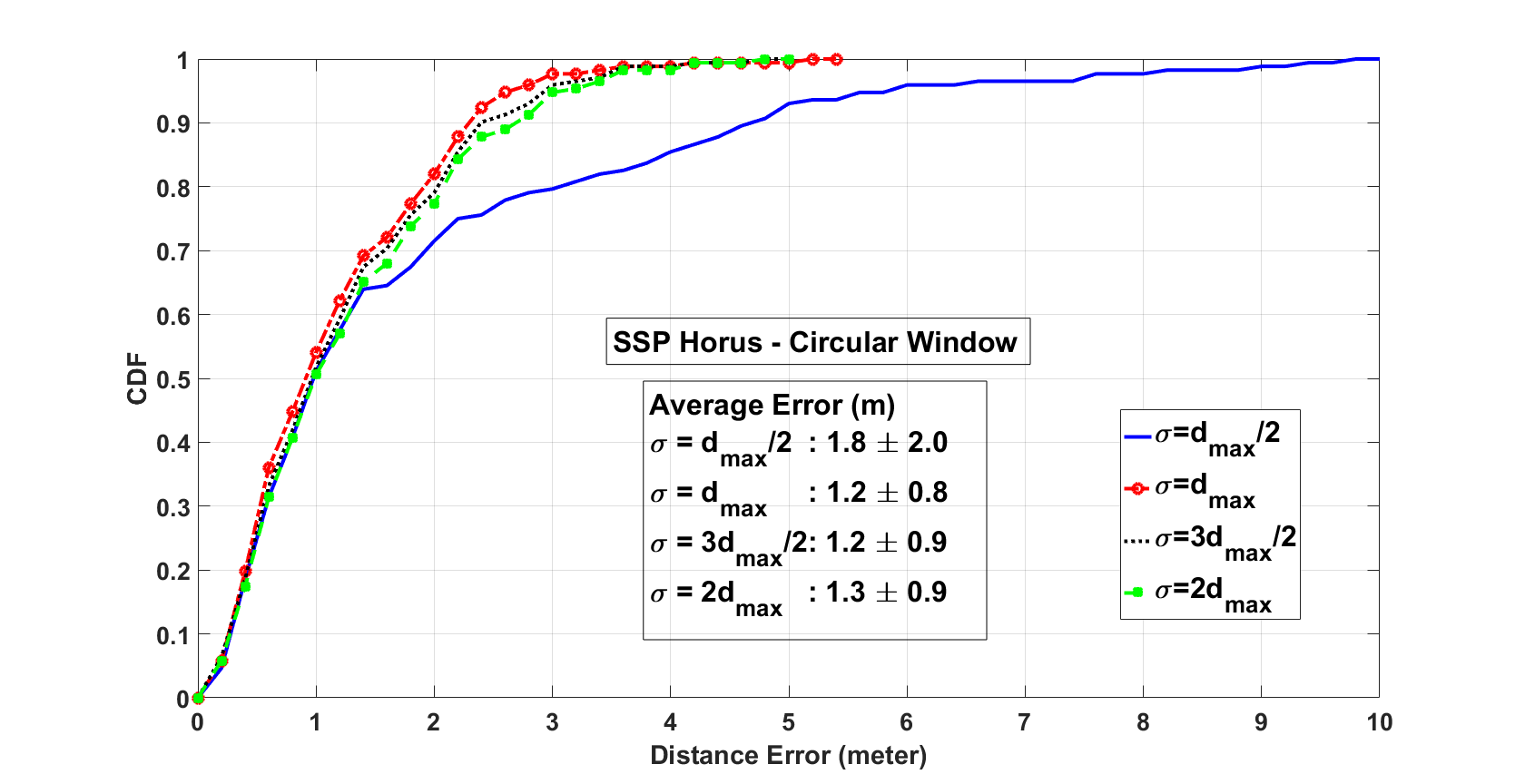}}\quad
\subfloat[\label{fig:Diff_Gaussian}]{\includegraphics[width=0.48\textwidth]{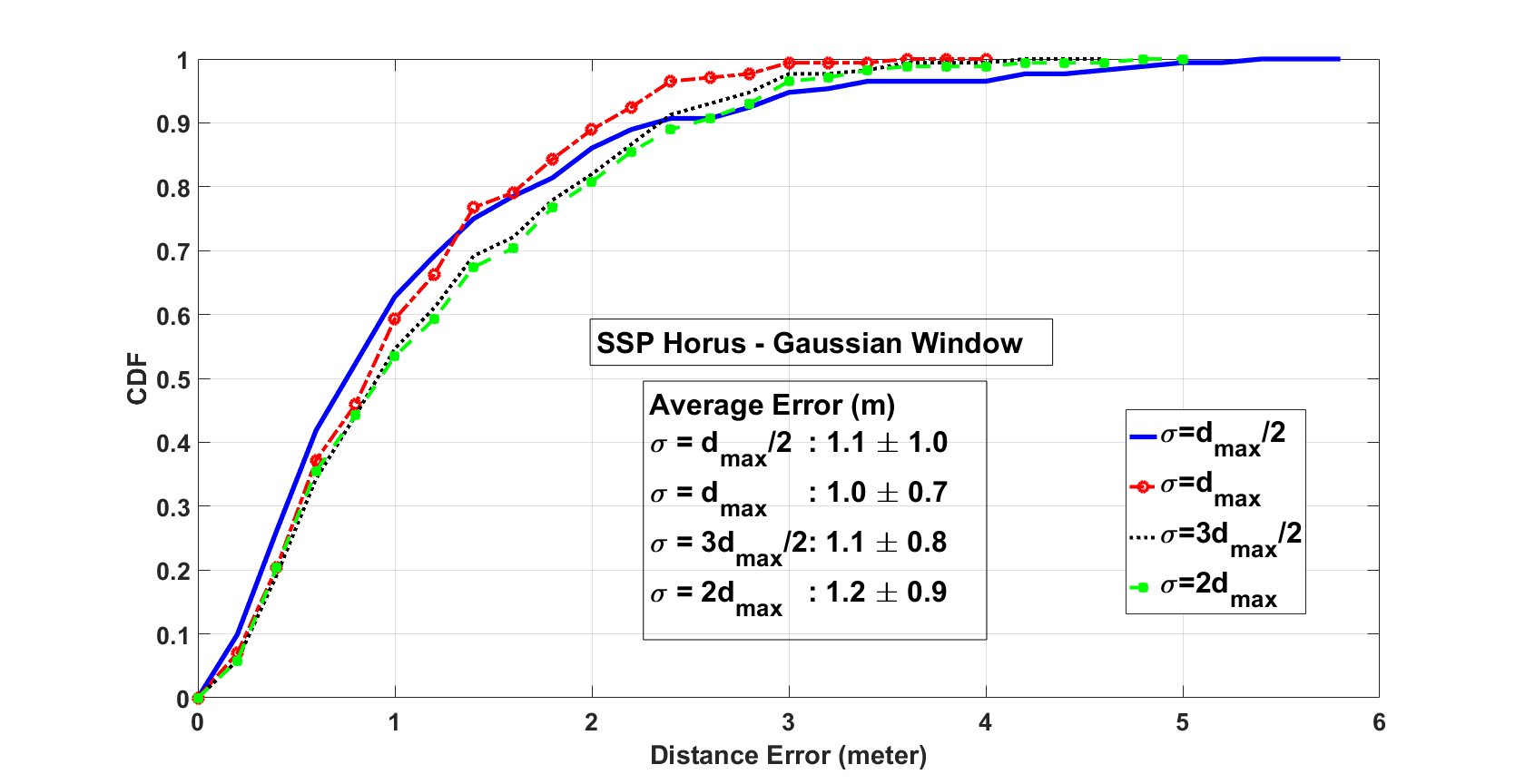}}\quad
\subfloat[\label{fig:Diff_Hann}]{\includegraphics[width=0.48\textwidth]{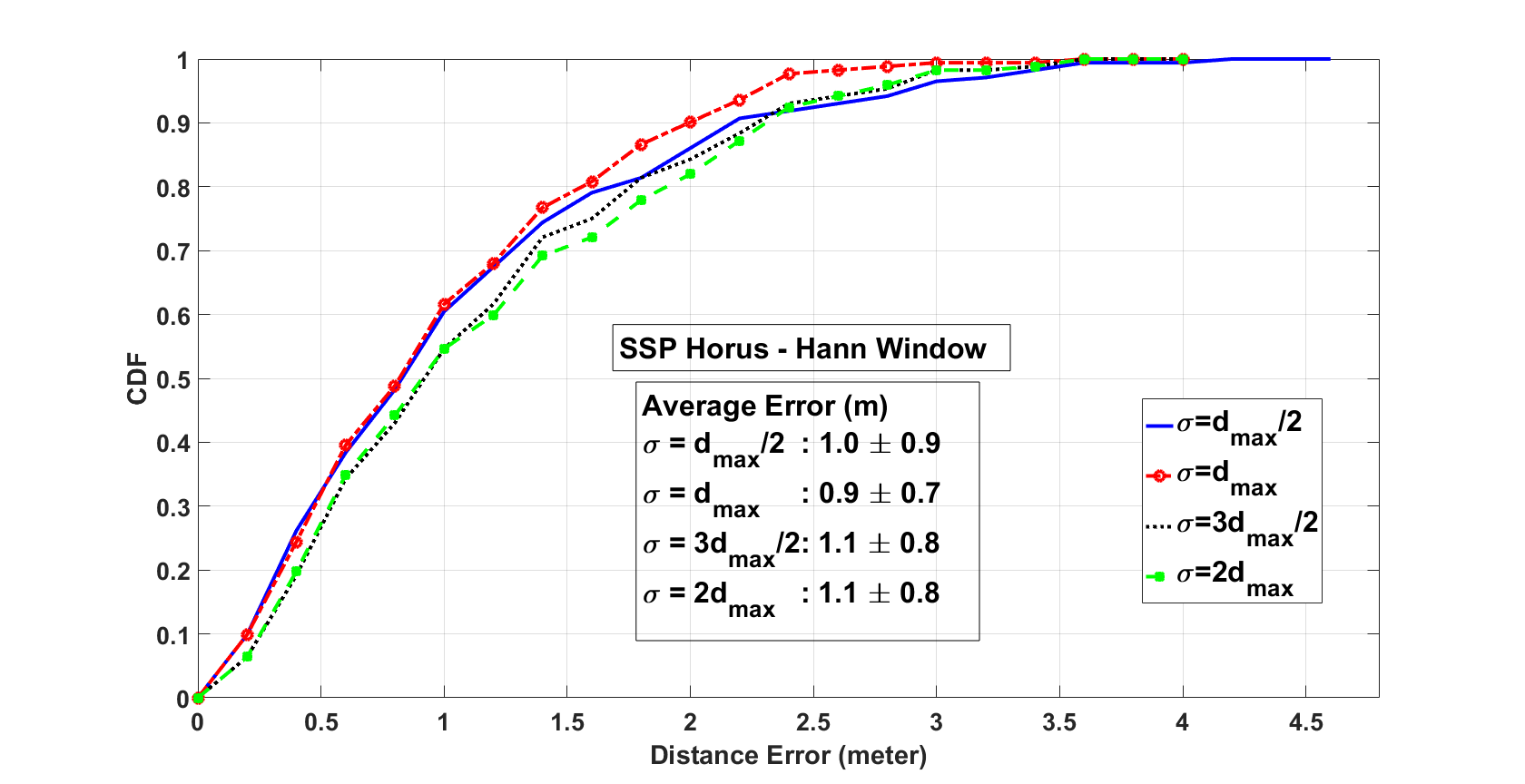}}\quad
\subfloat[\label{fig:Diff_Tukey}]{\includegraphics[width=0.48\textwidth]{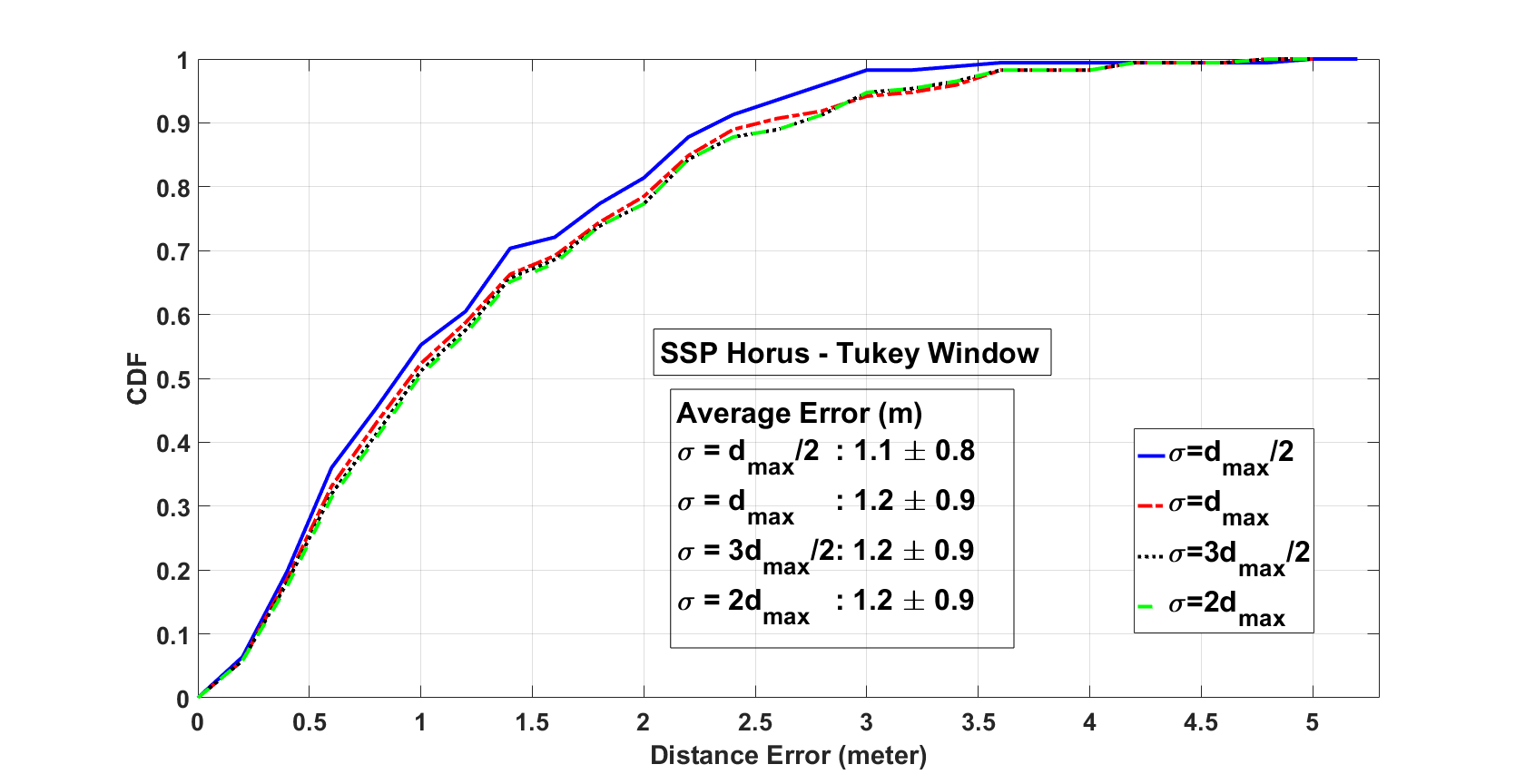}}
     \caption{CDF of the localization error of SSP Horus model with different window forms and $\sigma$ values (a) Circular window. (b) Gaussian window. (c) Hann window. (d) Tukey window.}
     \label{fig:Diff_Sigma_Horus}
\end{figure}

Fig.~\ref{fig:prob_heatmap} demonstrates the effects of various  SSP windows by their probabilistic heat maps with the color scale to the probability density. The red star in the map is the predicted previous location $\bm{l}_{prev}^{\prime}$. For each window we plotted both $\sigma=d_{max}$ and $\sigma=2d_{max}$. Clearly, the circular window only takes the area around the previous location with the highest probabilities (yellow color area) and eliminates the rest (blue color area). Gaussian window yields the most significant change when we increase the value of $\sigma$. When $\sigma=d_{max}$, a half of the area is ignored, while when $\sigma=2d_{max}$, most of the RPs are included with positive probabilities. Besides, Hann and Tukey have less dramatic changes than Gaussian when we increase $\sigma$. Furthermore, using Tukey window, the colors of different RPs are similar across the whole maps, indicating fewer differences among the probability densities of different RPs than the other window types. 
%%%%%%%%%%%%%%%%%%%%%%%%%%%%%%%%%%%%%%%%%%%%%%%%%%%%%%%%%%%%%%%%%%%%%%%%%%%%%%%%%%%%%%%%%%%%%%%%%%%%%%%%%%%%%%%
\subsubsection{Standard deviation $\sigma$}

 \begin{table}[!t]
\centering         
\caption{Average Errors of SSP Horus(meter)} 
\label{table:Diff_Sigma_Horus} 
\begin{tabular}{l c c c c} 
\hline           
\textbf{$\sigma$ (m)} & \textbf{Circular} & \textbf{Gaussian} & \textbf{Hann} & \textbf{Tukey}\\ 
$d_{max}/2$ &  $1.8\pm2.0$ & $1.1\pm1.0$ & $1.0\pm0.9$ & $1.1\pm0.8$\\
$d_{max}$ &  $1.2\pm0.8$ & $1.0\pm0.7$ & $0.9\pm0.7$ & $1.2\pm0.9$\\
$3d_{max}/2$&  $1.2\pm0.9$ & $1.1\pm0.8$ & $1.1\pm0.8$ & $1.2\pm0.9$\\
$2d_{max}$ &  $1.3\pm0.9$ & $1.2\pm0.9$ & $1.1\pm0.8$ & $1.2\pm0.9$\\
\hline         
\end{tabular} 
% \end{adjustbox}
\end{table}

\begin{table}[!t]
\centering         
\caption{Average Errors of SSP DGD(meter)} 
\label{table:Diff_Sigma_DGD} 
\begin{tabular}{l c c c c} 
\hline           
\textbf{$\sigma$ (m)} & \textbf{Circular} & \textbf{Gaussian} & \textbf{Hann} & \textbf{Tukey}\\ 
$d_{max}/2$  &  $1.8\pm2.4$ & $1.0\pm0.9$ & $1.1\pm0.9$ & $1.2\pm0.9$\\
$d_{max}$  &  $1.2\pm0.9$ & $1.1\pm0.8$ & $1.1\pm0.8$ & $1.1\pm0.9$\\
$3d_{max}/2$  &  $1.3\pm1.0$ & $1.1\pm0.9$ & $1.1\pm0.9$ & $1.3\pm1.0$\\
$2d_{max}$  &  $1.3\pm1.0$ & $1.2\pm1.0$ & $1.2\pm0.9$ & $1.3\pm1.1$\\
\hline         
\end{tabular} 
% \end{adjustbox}
\end{table}

\begin{table}[!t]
\centering         
\caption{Average Errors of SSP Kernel Method (meter)} 
\label{table:Diff_Sigma_Kernel} 
\begin{tabular}{l c c c c} 
\hline           
\textbf{$\sigma$ (m)} & \textbf{Circular} & \textbf{Gaussian} & \textbf{Hann} & \textbf{Tukey}\\ 
$d_{max}/2$  &  $1.2\pm1.0$ & 0.9$\pm0.8$ & $0.9 \pm 0.7$ & $1.0\pm0.8$\\
$d_{max}$  &  $1.2\pm0.9$ & $1.0\pm0.7$ & $1.0\pm0.8$ & $1.0\pm0.8$\\
$3d_{max}/2$  &  $1.1\pm0.8$ & $1.0\pm0.8$ & $1.0\pm0.8$ & $1.0\pm0.8$\\
$2d_{max}$  &  $1.1\pm0.9$ & $1.1\pm0.8$ & $1.1\pm0.8$ & $1.1\pm0.9$\\
\hline         
\end{tabular} 
% \end{adjustbox}
\end{table}

The width of the short term memory windows is determined by the standard deviation $\sigma$. Fig.~\ref{fig:window_form} shows that when $\sigma$ increases, the PDF spread out more, which means more RPs being included as candidate locations in SSP. If $\sigma$ is too small, many possible RP candidates will be ignored, leading to huge accumulated errors. On the other hand, if $\sigma$ is too large, ambiguous locations are introduced which severely affects the localization performance.

In order to study the impacts of $\sigma$ on SSP, Table~\ref{table:Diff_Sigma_Horus},~\ref{table:Diff_Sigma_DGD} and \ref{table:Diff_Sigma_Kernel} compare the average errors  with different $\sigma$ when SSP model is applied. For the circular window, all of the RPs having a distance bigger than $\sigma$ are not considered in the current location estimation. Therefore, a significant error is observed with a low $\sigma$ value, e.g., $1.8 \pm 2.0$ m at $\sigma=d_{max}/2$  with SSP Horus and SSP DGD. When $\sigma$ increases, the performance is improved with the average error being around $1.2\pm0.8$ m at $\sigma=d_{max}$. The lowest average error of $1.1\pm0.8$~m is reached when $\sigma=3d_{max}/2$. In contrast, the performance of the soft range windows such as Gaussian, Hann and Tukey, is insensitive to $\sigma$ value. When $\sigma=d_{max}$,  soft range windows with Horus and DGD  have the lowest average errors of $1.0\pm0.7$~m and $1.1\pm0.8$~m respectively. On the other hand, Kernel method reaches its best performance of $0.9\pm0.8$~m at $\sigma=d_{max}/2$. When $\sigma$ increases from $\sigma=d_{max}$ to $\sigma=2d_{max}$, there is only a slight change in the performance of all three types of soft range windows. The reason is that instead of a hard cutoff, soft range windows provide a smooth bias toward locations closer to the previous prediction while no locations are fully eliminated from consideration.         

Fig.~\ref{fig:Diff_Sigma_Horus} illustrates SSP Horus error CDFs with different $\sigma$ values. For the circular, Gaussian and Hann windows, when $\sigma=d_{max}/2$ the maximum errors of SSP Horus increase significantly to 10~m, 5.7~m and 4.5~m respectively. When $\sigma$ equals or is larger than $d_{max}$, all maximum errors are around 4-5~m. Especially, for Tukey window, the CDFs are similar even when $\sigma$ changes from $d_{max}/2$ to $2d_{max}$. 

In summary, $\sigma=d_{max}$ provides the best localization accuracy for most of the cases. Furthermore, Gaussian and Hann windows consistently show the comparable performance. Therefore, in the rest of the analysis, we choose Gaussian window at $\sigma=d_{max}$ as our SSP window.

%%%%%%%%%%%%%%%%%%%%%%%%%%%%%%%%%%%%%%%%%%%%%%%%%%%%%%%%%%%%%%%%%%%%%%%%%%%%%%%%%%%%%%%%%%%%%%%%%%%%%%%%%%%%%%%
\subsection{Performance Analysis}
\begin{figure}[!t]
\centering
\includegraphics[width=0.5\textwidth]{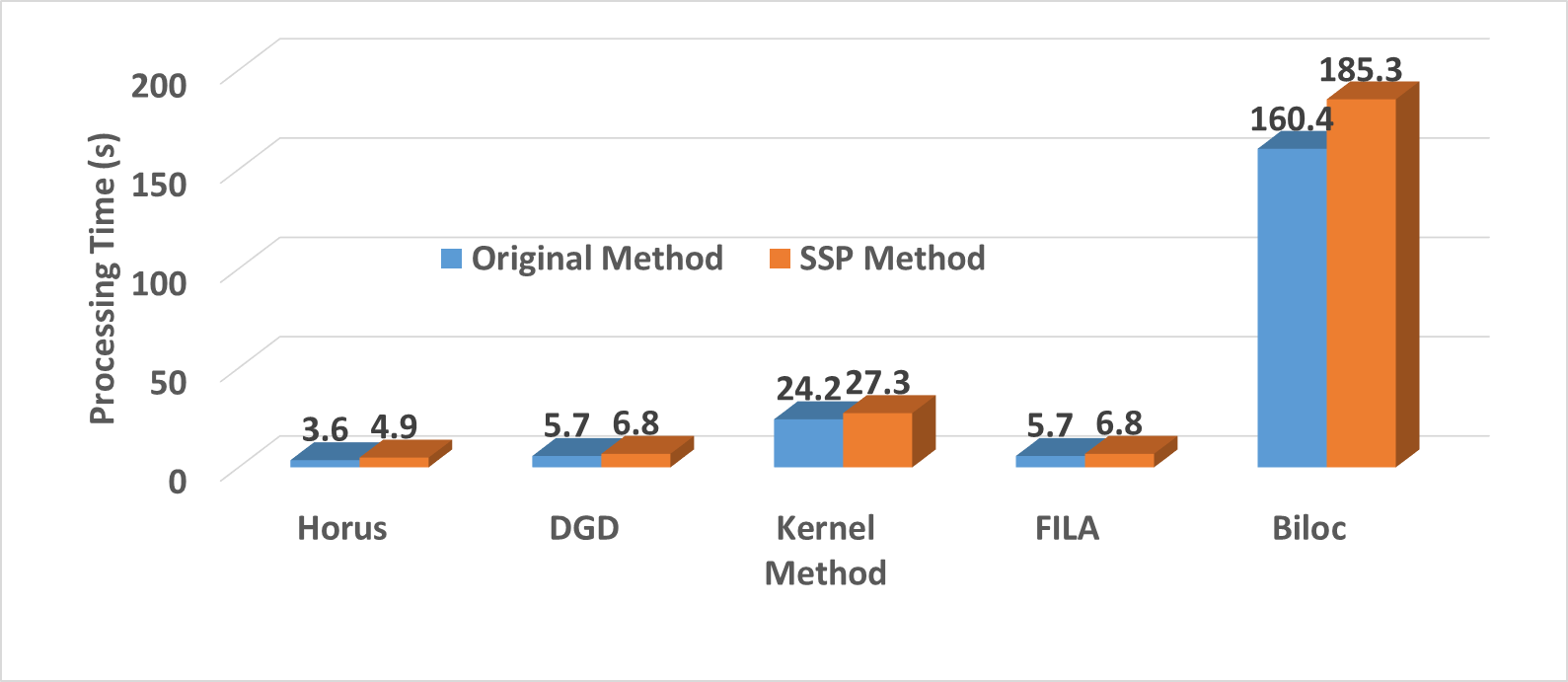}
\caption{Processing time of SSP methods compared with the conventional methods.}
\label{fig:processing_time}
\end{figure}

The conventional probabilistic approaches using Eq.~\eqref{prob2} have a complexity $O(N)$, where $N$ is the number of features as explained in Subsection~\ref{sec:Notation}. SSP adds a simple short term memory according to Eq.~\eqref{Gauss}, \eqref{Hann} and \eqref{Tukey} with no significant addition to the overall complexity. Fig.~\ref{fig:processing_time} shows the comparison of the processing time among all methods with and without SSP.  The computations are estimated on a Intel core i5-3320M 2.6 GHz based computer and an NVidia Geforce FTX 1050 GPU is used for training and testing neural network. For a fair comparison, the shown processing time here is for the testing phase. The processing time of SSP for all five implemented methods, including Horus, DGD, Kernel method, FILA and Biloc, is slightly higher than the original ones, ranging from $10 \%$ to $20 \%$ difference. In summary, the processing time of SSP model is comparable to that of the other conventional probabilistic approaches. 

\begin{figure}[!t]
     \centering
\subfloat[\label{fig:Horus_Error}]
{\includegraphics[width=0.48\textwidth]{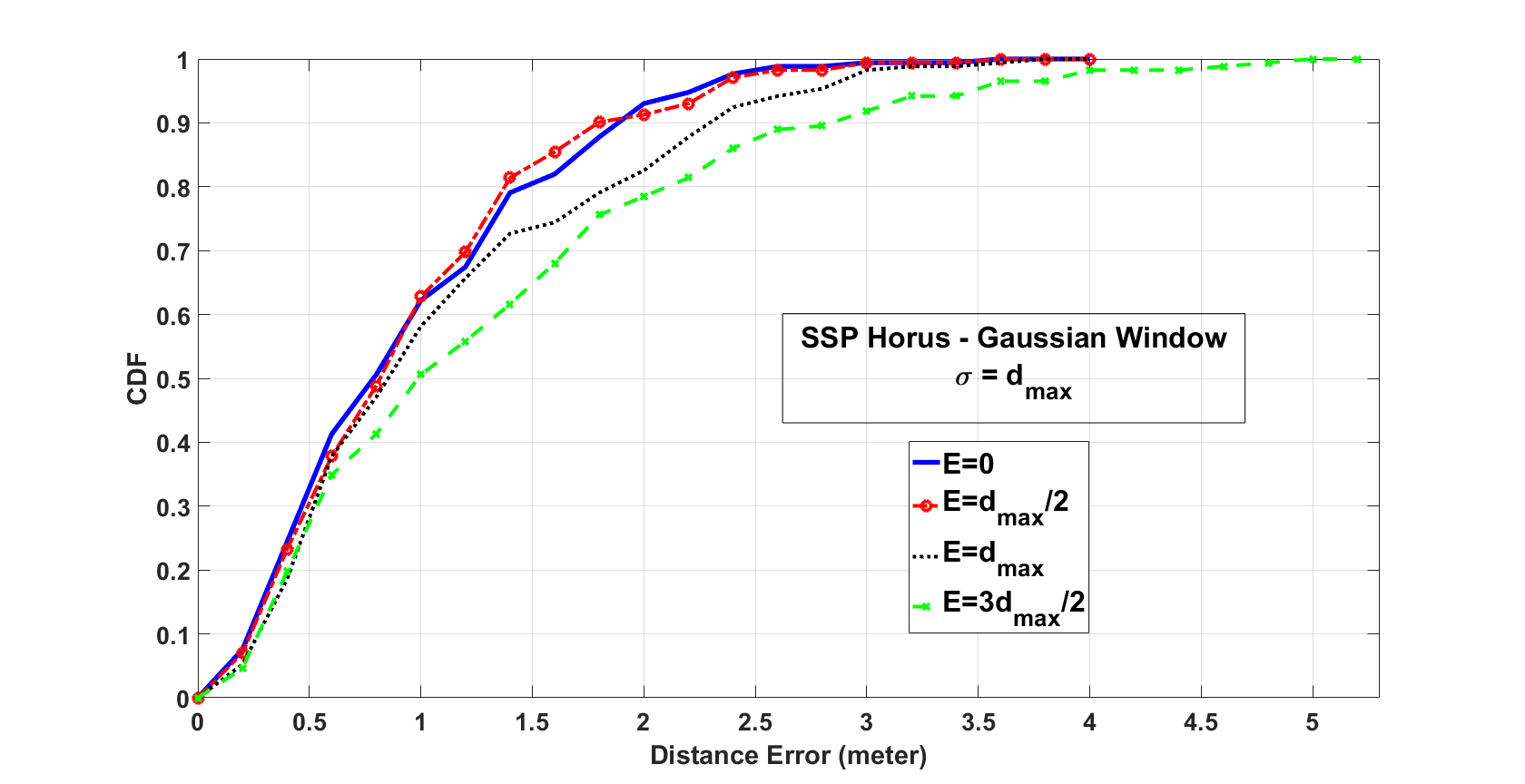}}\quad
\subfloat[\label{fig:DGD_Error}]{\includegraphics[width=0.48\textwidth]{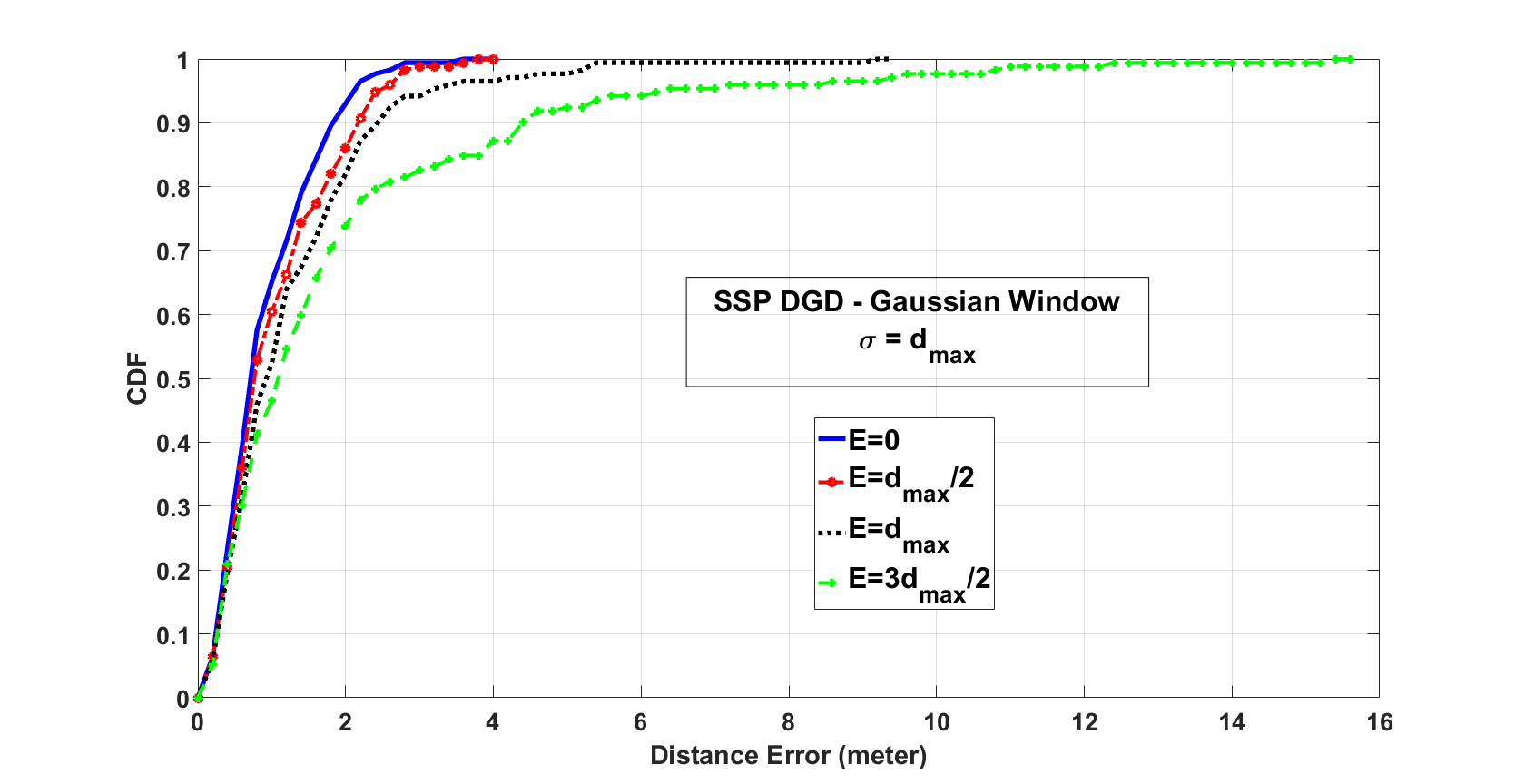}}\quad
\subfloat[\label{fig:Kernel_Error}]{\includegraphics[width=0.48\textwidth]{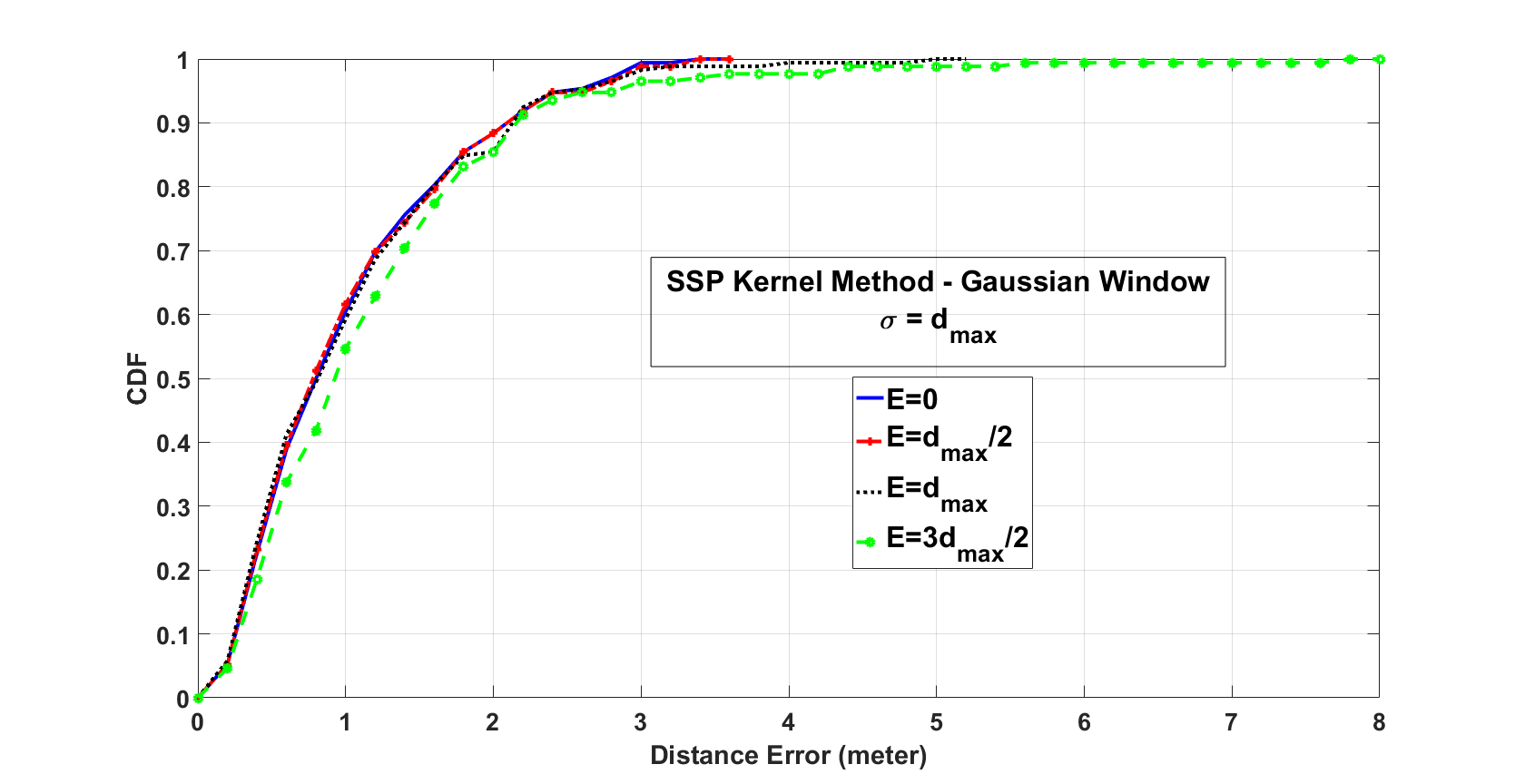}}
     \caption{CDF of localization errors of SSP based models using RSSI fingerprint in different error scenarios of historical data. (a) Horus. (b) DGD. (c) Kernel method.}
     \label{fig:CDF_Error}
\end{figure}

%\begin{figure}[!t]
%\centering
%\includegraphics[width=0.5\textwidth]{picture/ErrorAnalysis.pdf}
%\caption{CDF of localization errors of SSP based models using RSSI fingerprint in different error scenarios of historical data. (a) Horus. (b) DGD. (c) Kernel method.}
%\label{fig:CDF_Error}
%\end{figure}

Since SSP leverages the information of a user's previous position to estimate the current location, the performance of SSP depends on the accuracy of historical data. Note that all of our SSP results are based on the estimated history location with non-zero errors. In order to estimate the propagation error due to the imperfect prior location estimation, Fig.~\ref{fig:CDF_Error} illustrates the localization errors of SSP Kernel with both the ideal and erroneous history data. Starting with the perfect historical coordinate $\bm{h}(x,y)$ for every location in the testing trajectories mentioned in Section~\ref{sec:experiment}, an amount of error $E$ is added to $\bm{h}$. The erroneous prior location $\bm{h^{\prime}}(x^{\prime}, y^{\prime})$ is obtained as: $ x^{\prime} = x + x_{e} \, , \, y^{\prime} = y + y_{e}$, where $x_e$ and $y_e$ are random variables that follow Gaussian distribution 
\[ x_e \sim \mathcal{N}(0,\sigma_{x_{e}}^{2}) \, ; \, y_e \sim \mathcal{N}(0,\sigma_{y_{e}}^{2}) \, ; \, \sqrt{\sigma_{x_{e}}^{2}+\sigma_{y_{e}}^{2}} = E \] 

Fig.~\ref{fig:CDF_Error} shows the cases where $E$ is proportional to $d_{max}=4$~m. Clearly, if the error $E$ of the historical data is within $d_{max}/2$, the localization accuracy of all three methods are similar to the ideal case with a maximum error of ${\sim}4$~m. When $E$ increases to $d_{max}$, the accuracy of SSP Horus becomes slightly worse with $80\%$ of the error is under 2~m compared with 1.5~m of the ideal cases. On the other hand, SSP DGD and SSP Kernel are affected more significantly with the maximum error increasing to 9~m and 5~m respectively. As shown in Table~\ref{table:Diff_Sigma_Horus},~\ref{table:Diff_Sigma_DGD} and \ref{table:Diff_Sigma_Kernel},  average errors of all SSP models are around $d_{max}/4 \approx 1$~m, which indicates that SSP is robust to localization error of the previous position.  If the value of error $E$ is larger than $d_{max}$, i.e., $E=3d_{max}/2$, the performance will degrade and the accumulated errors start increasing. In order to solve the problem, the stationary time in a regular walking pattern can be utilized. According to the survey in Ref.~\cite{Wu2018}, the percentage of stationary time exceeds $80\%$ for most mobile users. During the no movement period, the number of RSSI readings collected in one-location ($S_{2}$) is sufficient to improve the conventional probabilistic model accuracy. Therefore, in order to enhance the accuracy when locating a user's position in a long trajectory, we can employ these stationary locations as alignment points where the prior locations can be ignored. In that case, the conventional probabilistic approaches can be exploited to estimate the user's location.

%%%%%%%%%%%%%%%%%%%%%%%%%%%%%%%%%%%%%%%%%%%%%%%%%%%%%%%%%%%%%%%%%%%%%%%%%%%%%%%%%%%%%%%%%%%%%%%%%%%%%%%%%%%%%%%
\section{Conclusions} \label{sec:conclude}
%%%%%%%%%%%%%%%%%%%%%%%%%%%%%%%%%%%%%%%%%%%%%%%%%%%%%%%%%%%%%%%%%%%%%%%%%%%%%%%%%%%%%%%%%%%%%%%%%%%%%%%%%%%%%%%
In conclusion, we have proposed a simple but efficient semi-sequential probabilistic model, which applies an additional short term memory step to enhance the performances of the indoor localization probabilistic approaches. This model leverages the information of the previous position to effectively determine the candidate location probability since the user's speed in an indoor environment is bounded. Three soft range windows including Gaussian, Hann and Tukey have been proposed as SSP models. Several experiments have been conducted including both RSSI and CSI fingerprints to demonstrate that SSP reduces the maximum error and boosts the performance of other existing probabilistic approaches by at least $25\% -30\%$. The error analysis also shows that the proposed SSP model is robust to localization error of the previous position.  

\bibliographystyle{IEEEtran}
\bibliography{Prob_Ref}

\end{document}